\documentclass[aps,prx,twocolumn,a4paper,superscriptaddress,10pt]{revtex4-2}

\usepackage{pdfpages}
\usepackage{etoolbox}
\apptocmd{\sloppy}{\hbadness 20000\relax}{}{}

\makeatletter
\AtBeginDocument{\let\LS@rot\@undefined}
\makeatother

\usepackage{epsfig}
\usepackage[makeroom]{cancel} 
\usepackage[normalem]{ulem} 
\usepackage{xcolor} 
\pagecolor{white}
\usepackage{amssymb,graphics,graphicx,amsfonts,amsmath,empheq}
\usepackage[breakable,most,many]{tcolorbox}
\usepackage{varwidth}
\usepackage{capt-of}
\usepackage[export]{adjustbox}
\usepackage{adjustbox}
\usepackage{natbib}
\usepackage{diagbox}
\usepackage{chngcntr}
\usepackage{enumerate}
\usepackage{physics}
\usepackage{bm}

\DeclareMathOperator\sgn{sgn}

\usepackage{hyperref}
\hypersetup{
    colorlinks,
    linkcolor={red!50!black},
    citecolor={blue!50!black},
    urlcolor={blue!80!black}}
    
\newcommand{\gp}{{\color{green!70!black}+}}
\newcommand{\gm}{{\color{green!70!black}-}}

\begin{document}

\title{Content-Addressable Memory with a Content-Free Energy Function}

\author{Félix Benoist}
\affiliation{Gulbenkian Institute of Molecular Medicine, Oeiras, Portugal}

\author{Luca Peliti}
\affiliation{Santa Marinella Research Institute, I-00058, Santa Marinella, Italy}

\author{Pablo Sartori}
\email{pablo.sartori@gimm.pt}
\affiliation{Gulbenkian Institute of Molecular Medicine, Oeiras, Portugal}

\begin{abstract}
Content-addressable memory, \textit{i.e.} stored information that can be retrieved from content-based cues, is key to computation. Besides natural and artificial neural networks, physical learning systems have recently been shown to have remarkable ability in this domain. While classical neural network models encode memories as energy minima, biochemical systems have been shown to be able to process information based on purely kinetic principles. This raises the question of whether neural networks can also encode information kinetically. Here, we propose a minimal model for content-addressable memory in which the kinetics, and not the energy function, are used to encode patterns. We find that the performance of this kinetic encoding is comparable to that of classical energy-based approaches. This highlights the fundamental significance of the kinetic stability of kinetic traps as an alternative to the thermodynamic stability of energy minima, offering new insights into the principles of computation in physical and synthetic systems.
\end{abstract}
\maketitle

\textbf{Introduction.} The ability to recover multiple memories from cues of their content is a fundamental computation performed by natural and artificial neural networks~\cite{Gerstner14,Hertz91,Ramsauer21}. The physical modeling of such a computational ability was pioneered by J.J. Hopfield in his celebrated 1982 article~\cite{Hopfield82}, which introduced an energy-based model for content-addressable memory. In this model, memories are modeled as patterns of activation of a fully-connected network of binary units representing neurons. The patterns are encoded via a set of energetic couplings and correspond to minima in the energy landscape of the network. Provided low crosstalk between patterns, each pattern can be retrieved via asynchronous updates from a content cue through equilibrium relaxation~\cite{Amit85,Hertz91}. 

In recent years, the performance of Hopfield networks has been significantly improved~\cite{Krotov23}, with modern versions showing promising results in machine learning~\cite{Liang22,Mehta19,Krotov21,Ramsauer21,Pham25}. More realistic networks of spiking neurons are also the subject of intense research in the context of biological neural computation~\cite{Podlaski24,Gerstner14,Rabinovich06}. Besides neurons, Hopfield-like encoding schemes have been used in models of biological liquid mixtures~\cite{Teixeira24}, multi-component self-assembly~\cite{Evans24,Murugan_PNAS15,Sartori20} and regulatory networks~\cite{Karin24}. 

\begin{figure}[!b]
	\centerline{\includegraphics[]{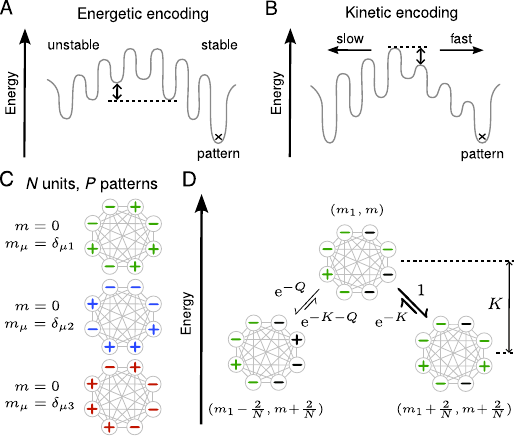}}
   \caption{\label{fig:sketch} {\it Schematics of kinetic encoding setup.} {\bf A.} In energetic encoding, the pattern state is preferred due to its lower energy. {\bf B.} In kinetic encoding, the pattern state is discriminated from other states with the same energy via the lower kinetic barriers along its pathway. {\bf C.} $N=8$ fully-connected units encode $P=3$ activity patterns. The retrieval of pattern $\mu=1$ corresponds to overlap values $(m_1,m_2,m_3)=(1,0,0)$ and activity $m=0$. {\bf D.} Sketch of the energy diagram for the retrieval of pattern $\mu=1$ from a state with only $-1$ errors (black minus signs) besides the cue (green plus and minus signs). Going down in energy, the system can either create a $+1$ error (left transition) or correct a $-1$ error (right transition), the latter being kinetically favored.}
\end{figure} 

Although there are many variants of Hopfield networks, patterns are always encoded as minima of an energy landscape [Fig.~\ref{fig:sketch}A]. Instead, we propose a fundamentally new encoding approach based on tuning the kinetics of the pathways leading to the patterns; see Fig.~\ref{fig:sketch}B. In this framework, each pattern becomes a kinetic trap, which can only be retrieved transiently. Therefore, patterns are not more stable than other states but are more kinetically accessible. Although such kinetic encoding has already been used to differentiate binding partners in copolymerization~\cite{Bennett79,Sartori13,Tsai06} and self-assembly processes~\cite{Benoist25}, in molecular error correction~\cite{Hopfield74,Sartori13,Ravasio24}, and in frenetic steering of random walkers~\cite{Lefebvre23}, it is so far absent from Hopfield neural networks. We anticipate that this kinetic approach results in a functional content-addressable memory network with capacity comparable to traditional Hopfield networks, along with unique features, such as glassy behavior~\cite{Bouchaud97,Bouchaud92}.
\newline

\textbf{Content-addressable memory.} Consider $N$ interacting binary units indexed $i=1,\dots,N$, which represent neurons in a neural network. Each unit can be active or inactive, which we characterize by its activity value, $\sigma_i=+1$ or $-1$, respectively. Among the $2^N$ system configurations, we select $P$ ``pattern states'' with random activities denoted $\xi_i^\mu$, $\mu=1,\dots,P$ and null mean; see Fig.~\ref{fig:sketch}C. We quantify similarity between an arbitrary state and pattern $\mu$ through the overlap $m_\mu=\sum_i\sigma_i\xi_i^\mu/N$, which generalizes the activity $m=\sum_i\sigma_i/N$.

We consider a dynamics satisfying detailed balance, such that the transition rate from state $\bm\sigma=(\sigma_1,\dots,\sigma_i,\dots,\sigma_N)$ to $\bm\sigma'=(\sigma_1,\dots,-\sigma_i,\dots,\sigma_N)$ is given by
\begin{equation}\label{eq:k_Glauber}
	k(\bm\sigma\to\bm\sigma') = \frac{\omega(\bm\sigma,\bm\sigma')}{1+\exp(\beta \Delta E(\bm\sigma,\bm\sigma'))}.
\end{equation}
Here, $\beta$ is the inverse temperature, $\omega(\bm\sigma,\bm\sigma')=\omega(\bm\sigma',\bm\sigma)$ is a bare kinetic rate, and $\Delta E=\mathcal H(\bm\sigma')-\mathcal H(\bm\sigma)$ is the energy difference, where $\mathcal{H}$ is the system's Hamiltonian.  Note that the energy differences and the bare kinetic rates have different symmetries: Whereas $\omega(\bm\sigma',\bm\sigma)$ is even in the exchange $\bm\sigma \leftrightarrow \bm\sigma'$, one has $\Delta E(\bm\sigma',\bm\sigma)=-\Delta E(\bm\sigma,\bm\sigma')$.

A model for content-addressable memory provides expressions for $\omega(\bm\sigma,\bm\sigma')$ and $\mathcal{H}(\bm\sigma)$ that ensure pattern retrieval. That is, if the system is initialized with a cue for pattern $\mu=1$, \textit{i.e.}, in a state for which $m_1$ is small but non-zero and $m_{\mu\neq 1} \ll m_1$, then the system should evolve over time toward pattern 1: $\bm\sigma=(\xi_1^1,\dots,\xi_N^1)$. In this pattern state, $m_1=1$ and $m_{\mu\neq1}\approx0$, due to the near-orthogonality of patterns for large $N$~\cite{Hertz91}.
\newline

\textbf{Energetic encoding.}  In classical Hopfield networks, neurons are coupled via a matrix $J_{ij}$ that encodes the $P$ patterns~\cite{Hopfield82,Hertz91}:
\begin{equation}\label{eq:Jij}
	J_{ij} = \frac{1}{N}\sum_{\mu=1}^P\xi_i^\mu\xi_j^\mu.
\end{equation}
The Hamiltonian of the network is then defined as $\mathcal H(\bm\sigma)=-\frac12\sum_{i,j\neq i}J_{ij}\sigma_i\sigma_j$. At low temperature, $\beta>1$, this setup ensures pattern retrieval provided that $P<P_{\rm max}$, with $P_{\rm max}\sim N$ [SM Sec.~\ref{sec:capa}]. Moreover, including higher-order couplings beyond the pairwise couplings of Eq.~\eqref{eq:Jij} leads to a dramatic increase in encoding capacity~\cite{Krotov16,Newman88,Demircigil17}.

Crucially, this encoding is such that, for $P<P_{\max}$, the patterns are minima of $\mathcal H$. Since the original Hopfield network considered the limit $\beta\to\infty$, there was no discussion of bare kinetic rates. Later works at finite temperature~\cite{Amit85,Kanter87}, as well as recent advances in energy-based models~\cite{Krotov16, Krotov21, Demircigil17}, all assume $\omega(\bm\sigma,\bm\sigma')=1$, so that all information encoding is energetic, \textit{i.e.} engraved in the Hamiltonian. 
\newline

{\bf Kinetic encoding.} We now provide an alternative approach to encoding patterns in Hopfield networks. We choose a Hamiltonian that is independent of the couplings $J_{ij}$ defined in Eq.~\eqref{eq:Jij}, but we let these couplings determine the bare kinetic parameters. Specifically, we set
\begin{align}\label{eq:H}
	\beta\mathcal H(\bm\sigma)= \frac{N}{2} K |m(\bm\sigma)|,
\end{align}
where $K\ge 0$ is the energetic drive in units of $\beta^{-1}$. The $2^{N/2}$ states with zero activity, $m=0$, are states of minimal energy, among which the $P$ patterns need to be discriminated. To this end, we
choose the bare kinetic parameters such that transitions that increase the overlap with the patterns are accelerated. In particular, we set
\begin{equation}\label{eq:omega_i}
\omega(\bm{\sigma},\bm{\sigma}')=\begin{cases}
	1,&\text{if }\ h_i\ge 0,\\
	\exp(-Q),&\text{if }\ h_i< 0,\end{cases}
\end{equation}
where $Q\ge 0$ is a discrimination barrier in units of $\beta^{-1}$, and $h_i(\bm{\sigma})=\sum_{j\neq i}J_{ij}\sigma_j$ is the effective field acting on unit $i$, which is the unit that changes between $\bm{\sigma}$ and $\bm{\sigma}'$. Since $h_i(\bm{\sigma})$ does not depend on $\sigma_i$, it follows that $\omega$ is symmetric with respect to the exchange $\bm\sigma\leftrightarrow\bm\sigma'$.

The choice of the expressions for $\mathcal{H}$ and $\omega$ made above can be justified as follows. Suppose that the system is initially set in a state such that a random fraction $m_1$ of units respects pattern $\mu=1$, while the rest is inactive ($\sigma_i=-1$), and so $m=m_1-1<0$ [SM Sec.~\ref{sec:theory}]. In this case, the energy function of Eq.~\eqref{eq:H} tends to increase activity toward $m=0$ by activating the inactive units indiscriminately. However, as per Eq.~\eqref{eq:omega_i}, the units with a positive local field will tend to activate first. Using Eq.~\eqref{eq:Jij}, we find that $h_i=\sum_\mu\xi_i^\mu m_\mu-\sigma_iP/N$, which, given $m_1\gg P/N$, approximates to $h_i\approx\xi_i^1m_1$. This implies that the first units to activate will be the ones for which $\xi_i^1=+1$. Therefore, we expect that the combination of these two effects results in activating the inactive units that are active in pattern 1; see Fig.~\ref{fig:sketch}D. The correction of these ``$-1$ errors'' ($\sigma_i=-1$ while $\xi_i^1=+1$) will lead to the simultaneous increase of $m_1$ and $m$ over time $t$. Importantly, the bare kinetic rates in Eq.~\eqref{eq:omega_i} are inherently unable to correct $+1$ errors, as they preferentially activate units with $\xi_i^1=+1$. In other words, we expect the present kinetic encoding model to perform pattern retrieval, provided that the non-cue part of the network is inactivated. We substantiate this claim below with extensive numerical simulations and analytical calculations.
\newline

\begin{figure}[t]
    \centerline{\includegraphics[]{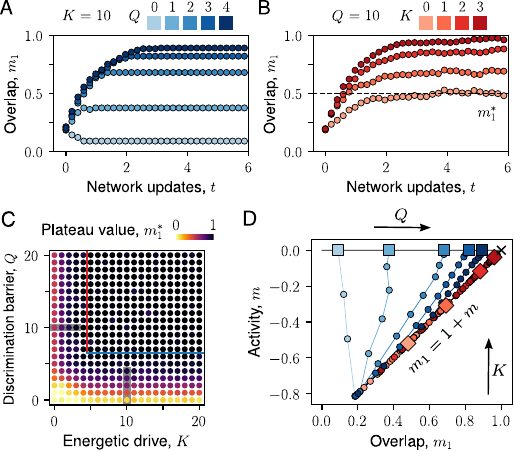}}
   \caption{\label{fig:param_sweep}{\it Successful retrieval of a single pattern.} {\bf A.} Stochastic trajectories for $P=1$ and $N=2^{10}$, starting from a random state with $m_1=0.2$ and $m=-0.8$; see [SM Sec.~\ref{sec:simu_details}] for simulation details. At a fixed value $K=10$, $m_1$ evolves toward a plateau whose value $m_1^*$ increases with $Q$. {\bf B.} At $Q=10$, the plateau value increases with $K$. The small fluctuations around the plateau value disappear with increasing $K$. {\bf C.} Phase diagram of pattern retrieval. The plateau value $m_1^*$ corresponds to a $1\%$ error for $Q$ and $K$ above the analytical estimates $\approx6.9$ and $4.6$ respectively (blue and red lines) [SM Sec.~\ref{sec:theory}]. The grey regions refer to the time trajectories shown in A-B. {\bf D.} The blue trajectories from {\bf A} have plateau values $m^*\approx0$ (larger squares), while for the red trajectories from B, $m_1^*\approx1+m^*$.}
\end{figure}

{\bf Retrieval of a single pattern.} As a starting point, we suppose that the couplings encode a single pattern ($P=1$). In this case, the network can be described by two quantities, namely the fractions of $-1$ errors and $+1$ errors, the latter referring to units with $\sigma_i=+1$ while $\xi_i^1=-1$ [SM Sec.~\ref{sec:theory}]. Based on the conceptual energy landscapes of Figs.~\ref{fig:sketch}B and D, we expect that retrieval of the pattern requires large values of the energetic drive $K$ and of the discrimination barrier $Q$. 
In fact, Fig.~\ref{fig:param_sweep} shows that this is the case, with $m_1$ reaching values close to 1 for high $K$ and $Q$. By contrast, if $Q$ is low (Fig.~\ref{fig:param_sweep}A), the descent in the energy function $\mathcal H$ (Eq.~\eqref{eq:H}) is unbiased, and thus the system gets trapped in an arbitrary state. On the other hand, if $K$ is low (Fig.~\ref{fig:param_sweep}B), the system will diffuse in a flat energy landscape, and the overlap saturates at an intermediate value. Therefore, both high $K$ and high $Q$ are necessary for pattern retrieval.

In [SM Sec.~\ref{sec:theory}], we analytically determine the conditions under which the system can retrieve a pattern from a cue. We find that for $K$ and $Q$ above certain thresholds $K_{\rm min}$ and $Q_{\rm min}$, the system successfully reaches the pattern state. Specifically, considering success as achieving less than $1\%$ error in the average overlap ($m_1=0.99$), in the thermodynamic limit $N\to\infty$, these thresholds are $K_{\rm min}\approx4.6$ and $Q_{\rm min}\approx6.9$; see Fig.~\ref{fig:param_sweep}C. 

To further investigate retrieval, we analyze trajectories of the overlap $m_1$ and activity $m$ over time. In the limit $K,Q\to\infty$, Fig.~\ref{fig:sketch}D shows that $m_1$ and $m$ increase simultaneously, such that $m_1(t)-m(t)$ is constant over time. Figure~\ref{fig:param_sweep}D displays examples of these trajectories, which, for increasing $K$ and $Q$, approach the predicted pathway to the pattern, $m_1(t)=1+m(t)$. As anticipated, to reach the pattern ($m_1=1$ and $m=0$), the initial state cannot contain any $+1$ errors since they cannot be corrected, \textit{i.e.}, $m(0)=m_1(0)-1$ [SM Sec.~\ref{sec:theory}].  
\newline

\begin{figure}[!t]
    \centerline{\includegraphics[]{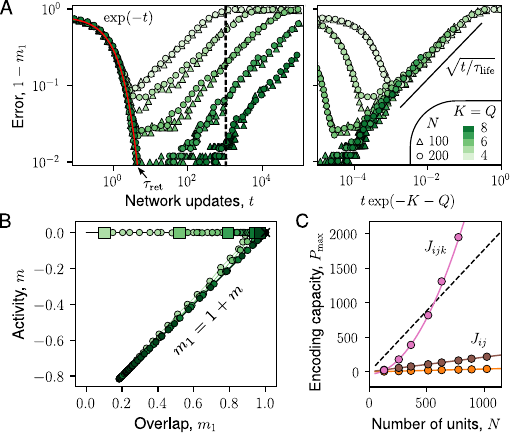}}
   \caption{\label{fig:stability}{\it Large kinetic stability and high encoding capacity.} {\bf A.} The system eventually escapes the kinetic trap ($m_1=1$) toward states with lower values of $m_1$. The retrieval dynamics follows the exponential prediction (red line). The escape dynamics rescaled horizontally by the lifetime, $t/\tau_{\rm life}$, is similar for different values of $K,Q,N$ with a slope of $\tfrac12$ in log-log (black line). Additionally, the minimum error decreases with increasing values of $K$ and $Q$. Here, $P=1$. {\bf B.} The trajectories in log time show that the eventual slow decrease of $m_1$ occurs near the $m=0$ line. The values at time $t=10^3$ (dashed line in {\bf A}) are shown as larger squares. {\bf C.} The plateau value of the overlap with the cued pattern, $m_1^*$, decreases with increasing number of patterns, $P$ [SM Sec.~\ref{sec:capa}]. For Hebbian pair couplings, the encoding capacity $P_{\rm max}$ at which $m_1^*=0.95$ scales with $N$. The capacity is higher for $m_1(0)=0.9$ (brown) than for $m_1(0)=0.2$ (orange), but still far from the Gardner bound~\cite{Gardner_JPhysA88_2,Shim93} (dashed line). By contrast, for three-body couplings, $P_{\rm max}$ scales with $N^2$ (pink).}
\end{figure}

{\bf Kinetic stability of the pattern.} Our model encodes patterns as kinetic traps from which the system eventually escapes. We quantify the kinetic stability of patterns by the ratio of their lifetime to their retrieval time. We obtain the retrieval time $\tau_{\rm ret}$, measured in number of network updates, by approximating the dynamics as follows. For $K,Q\to\infty$, retrieval reduces to finding the $-1$ errors to correct, so that the average error fraction $1-m_1$ decreases at a rate equal to its value [SM Sec.~\ref{sec:theory}]. Therefore, in the thermodynamic limit $N\to\infty$, the error fraction decays as $\exp(-t)$. It follows that $\tau_{\rm ret}$ is independent of $K,Q$ and $N$. Defining $\tau_{\rm ret}$ as the time when $m_1$ reaches $0.99$ starting from $m_1(0)=0.2$, we find $\tau_{\rm ret}\approx4.4$, consistent with the simulation data in Fig.~\ref{fig:stability}A. 

After pattern retrieval ($m_1\approx1$ and $m\approx0$), the system remains near the energy minimum $m=0$, but slowly drifts away from the pattern state, as reflected by the slow decrease of $m_1$ in Fig.~\ref{fig:stability}A-B. This drift arises because states with $m=0$ but lower $m_1$ are more numerous and thus entropically favored. The decrease in $m_1$ proceeds via double errors, \emph{i.e.}, deactivation of a unit and subsequent activation of another, to keep the overall network activity close to $0$. These double errors occur at a rate $\sim\exp(-K)\exp(-Q)$, as sketched in Fig.~\ref{fig:sketch}D.

This escape dynamics is more complex than the retrieval dynamics, but we can estimate the lifetime of patterns as the time to make a fraction of $N$ double errors, which gives a number of network updates $\tau_{\rm life}\sim \exp(K+Q)$ [SM Sec.~\ref{sec:theory}]. Therefore, the ratio of lifetime to retrieval time is 
\begin{equation}\label{eq:nretlife}
	\tau_{\rm life}/\tau_{\rm ret}\sim \exp(K+Q),
\end{equation}
and so large values of $K$ and $Q$ result in long-lived patterns relative to their retrieval time, as shown in Fig.~\ref{fig:stability}A. We observe that the eventual error increases approximately as $1-m_1\sim\sqrt{t/\tau_{\rm life}}$, a behavior for which we have not yet been able to provide an explanation. (A diffusive behavior would yield a square-root dependence not for the average error but for the fluctuations around that average.) To conclude, the kinetic encoding approach in Eqs.~(\ref{eq:H}-\ref{eq:omega_i}) enables fast retrieval of a single pattern for strong discrimination and large driving.
\newline

{\bf Encoding capacity.} We now turn to the case where the couplings encode multiple patterns ($P>1$). In this scenario, errors arise from crosstalk between patterns, which become increasingly prevalent as $P$ grows. For the same large positive values of $K$ and $Q$ that lead to retrieval for $P=1$, if the number of stored patterns exceeds a certain maximum, $P > P_{\rm max}$, retrieval fails due to crosstalk. We find that at low temperature, encoding capacity $P_{\rm max}$ scales linearly with the system size $N$ as in the classical case of energetic encoding~\cite{Hertz91}; see Fig.~\ref{fig:stability}C. This is expected, since kinetic encoding relies on the same local fields $h_i$ and is therefore subject to the same crosstalk constraints. At finite temperature, $P_{\rm max}$ additionally depends on $K$ and $Q$ [SM Sec.~\ref{sec:phases}].

\begin{figure}[!t]
    \centerline{\includegraphics[]{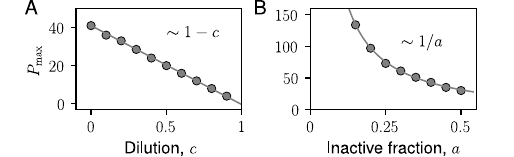}}
   \caption{\label{fig:dilute_sparse} {\it Dilute couplings and sparse patterns.} {\bf A.} For dilute connections between units, the encoding capacity $P_{\rm max}$ decreases linearly with the dilution. {\bf B.} For patterns with a vanishing fraction of inactive units, $P_{\rm max}$ diverges.}
\end{figure}

To better situate our work within current research, we study three simple model variants. First, we replace the pair couplings of Eq.~\eqref{eq:Jij} by three-body couplings $J_{ijk}=\sum_\mu\xi_i^\mu\xi_j^\mu\xi_k^\mu/N^2$ and the local field in Eq.~\eqref{eq:omega_i} by $h_i(\bm{\sigma})=\sum_{j,k\neq i}J_{ijk}\sigma_j\sigma_k$. This leads to an encoding capacity $P_{\rm max}\sim N^2$ [Fig.~\ref{fig:stability}C], in agreement with predictions from energetic encoding in modern Hopfield networks~\cite{Krotov16,Demircigil17,Newman88}. Second, motivated by biological plausibility, we examine the impact of diluted connectivity by pruning a fraction $c$ of the pair couplings~\cite{Hertz91}. As expected, $P_{\rm max}$ decreases as dilution increases; see Fig.~\ref{fig:dilute_sparse}A and [SM Sec.~\ref{sec:variants}]. Finally, we consider patterns with a fraction $a\le\frac12$ of inactive units ($\xi_i^\mu=-1$). Consistent with classical results on energetic encoding~\cite{Tsodyks88,Okada96}, we find that the capacity diverges in the sparse limit $a\to0$; see Fig.~\ref{fig:dilute_sparse}B and [SM Sec.~\ref{sec:variants}]. Together, these results underscore the versatility of kinetic encoding.
\newline

\begin{figure}[!t]
   \centerline{\includegraphics[width=.87\linewidth]{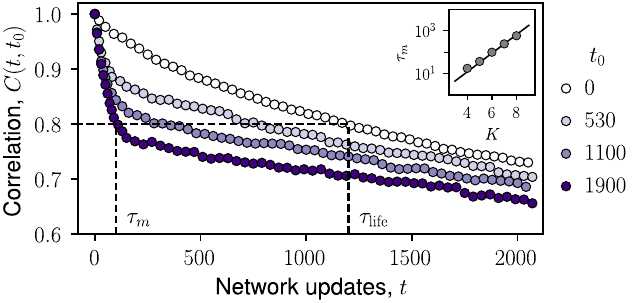}}
   \caption{\label{fig:coor_drop}{\it Temporal correlations are not time-translation invariant.} Beginning from the pattern state at $t=0$, the correlation between states at times $t_0$ and $t+t_0$ depends on the waiting time $t_0$. For $t_0\gtrsim \tau_{\rm life}$, $C(t,t_0)$ abruptly drops for $t\sim \tau_m$, whereas $C(t,0)$ decreases over timescale $\tau_{\rm life}\gg \tau_m$. Here, $N=100$, $P=1$, $m_1(0)=1$ and $K=Q=6$ such that the lifetime at which $m_1=0.8$ is $\tau_{\rm life}\approx1200$, while $C(\tau_m,1900)=0.8$ for $\tau_m\approx 100$. \textit{Inset:} Timescale $\tau_m$ increases exponentially with the energetic drive, $K$.} 
\end{figure}

{\bf Ageing relaxation.} As we have shown, the main differences between kinetic and energetic Hopfield networks lie in the dynamics, $\bm\sigma(t)$, not the capacity. To probe dynamics further, we now track time correlations
\begin{equation}\label{eq:Cnn0}
	C(t,t_0)=\frac1N\sum_i\sigma_i(t+t_0)\sigma_i(t_0),
\end{equation}
during escape from a single encoded pattern, setting $\bm\sigma(0)=(\xi_1^1,\dots,\xi_N^1)$. Figure~\ref{fig:coor_drop} shows that $C(t,t_0)$ depends strongly on the waiting time $t_0$. This absence of time-translation invariance is typical of glassy systems with widely separated timescales~\cite{Bouchaud97,Bouchaud92}. Importantly, unlike typical ageing glasses whose relaxation slows as the waiting time $t_0$ increases~\cite{Berthier02}, here the relaxation accelerates with increasing $t_0$. This behavior arises from an initial drop in the correlation for $t_0\gtrsim \tau_{\rm life}$ and $t\sim \tau_m$, the timescale over which errors move between units [SM Sec.~\ref{sec:ageing}]. The additional timescale $\tau_m\sim\exp(K)$ (see inset) lies between the retrieval time and the pattern lifetime, introducing an additional layer of complexity to the network dynamics.
\newline

{\bf Extension to continuous units.} So far, we have restricted units to binary values. To test the generality of our kinetic encoding framework, we extend it to networks of continuous units~\cite{Hopfield84, Behera23, Herron23}, which are not only more biologically plausible but also form the basis of dense associative memory models~\citep{Lucibello24, Krotov21}. For simplicity, each unit has a sigmoidal response $g(x_i)\in(-1,1)$ to input $x_i\in\mathbb R$ with gain $\lambda$. We consider synchronous dynamics in the form of a noisy gradient descent in an energy landscape $E(\mathbf x)$:
\begin{align}\label{eq:dxdt_gen}
    \frac{\text d x_i}{\text d t}=-\mu_i(\mathbf x)\frac{\partial E}{\partial x_i} + \sqrt{2D_i(\mathbf x)}\eta_i(t),
\end{align}
where $\eta_i$ are independent Gaussian white noises. Equilibrium dynamics then requires that the diffusivities and mobilities be linked as $D_i=\mu_i/\beta$, and that $\partial_j\partial_i\ln\mu_i=\partial_i\partial_j\ln\mu_j$~\cite{van_Kampen07}[SM Sec.~\ref{sec:variants}]. For instance, the classical continuous-unit Hopfield model~\cite{Hopfield84} corresponds to a noiseless version of Eq.~\eqref{eq:dxdt_gen}, where $E(\mathbf x)$ is function of the couplings $J_{ij}$ of Eq.~\eqref{eq:Jij}, while $\mu_i=1/g'(x_i)$ carries no pattern information.

For kinetic encoding, we instead propose an energy function free of the $J_{ij}$ matrix:
\begin{equation}\label{eq:E(x)_kin}
	E(\mathbf x)=\frac K\lambda|m(\mathbf x)| + \sum_i\int_0^{g(x_i)}g^{-1}(y)\text dy,
\end{equation}
where the first term echoes Eq.~\eqref{eq:H} with activity $m=\sum_ig(x_i)/N$ and energetic drive $K$, and the second term bounds the output dynamics. Pattern information enters solely through unit-specific frequencies $\omega_i(\mathbf x)$ equal to the bare kinetic rates of Eq.~\eqref{eq:omega_i}, with local fields $h_i(\mathbf x)=\sum_{j\neq i}J_{ij}g(x_j)$ and discrimination barrier $Q$. Incorporating these frequencies into the mobilities and diffusivities as $\mu_i=\omega_i(\mathbf x)/g'(x_i)$ and $D_i=\mu_i/\beta$ ensures detailed balance, since $h_i$ is independent of $x_i$ [SM Sec.~\ref{sec:variants}]. The resulting equilibrium dynamics are 
\begin{equation}\label{eq:dxdt_kin}
	\frac{\text dx_i}{\text dt}=\omega_i(\mathbf x)\left[-\frac K\lambda\sgn(m(\mathbf x))-x_i\right]+\sqrt{2D_i(\mathbf x)}\eta_i(t).
\end{equation}

As in the discrete case, in a network cued with pattern $\mu=1$ and negative activity, $m<0$, we expect the $-1$ errors (units with $g(x_i)<0$ while $\xi_i^1=+1$) to activate first, due to their higher update frequencies, $\omega_i=1$. Their greater mobility drives them to descend the energy gradient faster than the slower units with $\omega_i=\exp(-Q)$, channeling the network toward the pattern state. 

In [SM Sec.~\ref{sec:variants}], we investigate Eq.~\eqref{eq:dxdt_kin} without noise and find that the model reproduces the key features of the discrete case: transient retrieval, kinetic stability of patterns, and extensive capacity. The main differences concern timescales: the lifetime of patterns no longer depends on $K$, while the retrieval time decreases slightly with increasing $K$. The resulting simultaneous increase in speed and accuracy is typical of kinetic encoding~\citep{Bennett79,Sartori13,Benoist25}. 
\newline

{\bf Discussion.} In classical neural network models of content-addressable memory, patterns are stored as distinct energy minima. Our novel kinetic encoding approach instead sculpts the pathways leading to pattern states, enabling their reliable discrimination among a sea of erroneous energy minima. Through both simulation and theory, we have shown that our networks can retrieve many patterns transiently yet rapidly. Similar transient-retrieval phenomena have recently been reported in energy-based models~\cite{Clark25,Nicoletti25} and in transformer architectures~\cite{D'Amico24}. Our model however presents an important limitation as retrieval is only possible if the non-cue part of the pattern is inactive. We expect that this limitation can be circumvented by appropriate use of neural activation thresholds. The four extensions considered (three-body couplings, dilute connectivity, sparse patterns and continuous units) highlight the relevance of our work to neuroscience and artificial neural networks [SM Secs.~\ref{sec:capa} and~\ref{sec:variants}]. In future work, we may attempt to derive the dependence of $P_{\rm max}$ on $K$ and $Q$ (\textit{e.g.} using dynamical mean-field theory~\cite{Sompolinsky88,Clark24}) by tackling the complex noise structure in the continuous case.

This work builds on a previous study of kinetic encoding in polymerizing strings, where we similarly observed the rapid retrieval of multiple heteromeric target structures from the same components~\cite{Benoist25}. Therefore, we expect that the design principles outlined here may extend to other memory retrieval systems, in particular physical systems where detailed balance plays a central role~\cite{Dunn15,Yan17,Teixeira24}. Beyond the retrieval of static (or evolving~\cite{Schnaack22}) patterns, kinetic encoding also holds potential for applications in the retrieval of memory sequences~\cite{Sompolinsky86,Seliger03,Herron23} -- an area of research that has become particularly active in the context of learning within physical systems~\cite{Stern20,Keim19,Mandal24,Osat23}.
\newline

{\bf Acknowledgments.} This work was partly supported by a laCaixa Foundation grant (LCF/BQ/PI21/11830032) and core funding from the Gulbenkian Foundation.


%

\newpage

$ $

\newpage
\widetext
\begin{center}
\textbf{\Large Supplemental Material for ``Content-Addressable Memory with a \\\vspace{1.5mm} Content-Free Energy Function''}
\end{center}
\setcounter{section}{0}
\setcounter{equation}{0}
\setcounter{figure}{0}
\setcounter{table}{0}
\setcounter{page}{1}
\makeatletter
\renewcommand{\theequation}{S\arabic{equation}}
\renewcommand{\thefigure}{S\arabic{figure}}
\renewcommand{\thesection}{S\arabic{section}}
\vspace{5mm}

\section{Low-temperature encoding capacity}\label{sec:capa} 
Here, we assess the relevance of our work to artificial neural networks by systematically comparing the encoding capacity of our model with three benchmarks: classical Hopfield networks, the Gardner bound, and modern Hopfield networks. Our analysis shows that the performance of our model matches that of state-of-the-art content-addressable memory architectures. This corresponds to a discussion of Fig.~\ref{fig:stability}C.
\newline

{\bf Classical Hopfield networks.}
The zero-temperature encoding capacity of energetic encoding can be estimated from the stability of pattern states as follows ~\cite{Hertz91}. We focus on the case where the transition rate from state $\bm\sigma=(\sigma_1,\dots,\sigma_i,\dots,\sigma_N)$ to $\bm\sigma'=(\sigma_1,\dots,-\sigma_i,\dots,\sigma_N)$ is given by Eq.~\eqref{eq:k_Glauber} with $\Delta E(\bm\sigma,\bm\sigma')=2h_i\sigma_i$ and $\omega(\bm\sigma,\bm\sigma')=1$. At zero temperature, \textit{i.e.} $\beta\to\infty$, the dynamics reduces to a straightforward update rule for the activity $\sigma_i$ of each unit $i$:
\begin{equation}\label{eq:update_ener}
	\sigma_i = \sgn h_i = \sgn\Big(\sum_{j\neq i}J_{ij}\sigma_j\Big),
\end{equation}
where $J_{ij} = \sum_\mu\xi_i^\mu\xi_j^\mu/N$ [Eq.~\eqref{eq:Jij}]. The stability of pattern $\mu=1$ depends on its crosstalk with the remaining patterns, which we evaluate via $C_i^1 = \sum_{j\neq i}\sum_{\mu>1}\xi_i^\mu\xi_j^\mu\xi_j^1\xi_i^1/N$. In the limit $N,P\to\infty$ and for $\bm\sigma=(\xi_1^1,\dots,\xi_N^1)$, we decompose the local field as
\begin{equation}
	h_i\to \xi_i^1 + \frac1N\sum_{j\neq i}\sum_{\mu>1}\xi_i^\mu\xi_j^\mu\xi_j^1= \xi_i^1(1+C_i^1).
\end{equation}
Therefore, the update rule $\sigma_i=\xi_i^1\sgn(1+C_i^1)$ leads to the stability of pattern $1$ unless the crosstalk term $C_i^1$ is smaller than $-1$. Given random patterns with equal probabilities of $\xi_i^\mu=\pm1$, $C_i^1$ is $1/N$ times the sum of $NP$ random variables with values $\pm1$. It thus has zero mean and variance $\alpha=P/N$, which gives the rudimentary zero-temperature stability condition $\alpha<1$, \textit{i.e.}, 
\begin{equation}
	P<N. 
\end{equation}
More precise calculations give stability thresholds on the order of $\alpha_c\approx0.14$, \textit{i.e.} $P_c\approx0.14N$~\cite{Hertz91,Amit85}. However, for such values of $\alpha$, the pattern states correspond only to local minima, which are metastable. They instead become global minima for values of $\alpha$ smaller than $\alpha_\text{max}\approx0.05$, \textit{i.e.} $P_\text{max}\approx0.05N$. Thus, pair interactions can encode a number of patterns proportional to the system size.
\newline

{\bf Gardner bound.}
Beyond the particular Hebb rule for the couplings $J_{ij}$, one can ask a broader question: What are the optimal couplings for encoding a given set of $P$ patterns? This optimization problem is challenging, both numerically and analytically. A closely related question concerns the maximum number of patterns that can be stored with a non-zero basin of attraction, given optimal binary couplings $J_{ij}\in\{+1,-1\}$. This problem was tackled by E. Gardner and collaborators using replica calculations~\cite{Gardner_JPhysA88_2}. For uncorrelated patterns with zero mean activity and zero-error recall at zero temperature, they obtained the so-called Gardner bound $\alpha_G=4/\pi$, which largely exceeds the stability threshold for Hebbian couplings $\alpha_c\approx0.14$. The calculation was later extended to the case of a finite overlap threshold $m_1^{\rm th}\le1$~\cite{Shim93}. In this case, the maximum capacity reads 
\begin{equation}\label{eq:alpha_G}
	\alpha_G(m_1^{\rm th})=\frac2\pi\left(\int_0^T\frac{\text dt}{\sqrt{2\pi}}\exp(-t^2/2) t^2\right)^{-1}, \qq{with} T=\sqrt2\erf^{-1}(m_1^{\rm th}),
\end{equation}   
which gives $\alpha_G\approx1.8$ for $5\%$ error. We note, however, that a key advantage of Hebbian learning lies not only in its large capacity but also in its ease of training. By contrast, attaining the Gardner bound requires complex nonlinear optimization, which has limited practical relevance for current research in artificial neural networks.
\newline 

{\bf Modern Hopfield networks.}
Including couplings beyond pairwise has been shown to drastically increase encoding capacity~\cite{Krotov16,Newman88,Demircigil17}. For instance, three-body couplings that lead to pattern retrieval are given by $J_{ijk}=\sum_\mu\xi_i^\mu\xi_j^\mu\xi_k^\mu/N^2$, with a local field $h_i(\bm\sigma)=\sum_{j,k\neq i}J_{ijk}\sigma_j\sigma_k$. Following the same argument as above, the local field reads $h_i\to\xi_i^1(1+D_i^1)$, with a crosstalk term $D_i^1 = \sum_{j,k\neq i}\sum_{\mu>1}\xi_i^\mu\xi_j^\mu\xi_k^\mu\xi_k^1\xi_j^1\xi_i^1/N^2$. This results in the stability condition $P<N^2$. More generally, for $n-$body interactions, the zero-temperature encoding capacity scales as $P_{\rm max}\sim N^{n-1}$, with exponential scaling in the limit $n\to\infty$~\cite{Demircigil17,Lucibello24}. Most current research on artificial neural networks focuses on such architectures, with huge potential in topics such as large language models and transformer architectures~\cite{Ramsauer21}. We moreover note that many applications of modern Hopfield networks occur in the regime $P>P_{\rm max}$, in which spurious states are desirable, as they correspond to information generation~\cite{Pham25}.
\newline

{\bf Kinetic encoding.}
For kinetic encoding, the transition rate $k(\bm\sigma\to\bm\sigma')$ from Eq.~\eqref{eq:k_Glauber} has parameters 
\begin{equation}
\beta\Delta E(\bm\sigma,\bm\sigma')=\begin{cases}
	-K\sigma_i\sgn m,&\text{if }\ m\neq 0,\\
	K,&\text{if }\ m=0,\end{cases}
\end{equation}
and $\omega(\bm\sigma,\bm\sigma')$ as in Eq.~\eqref{eq:omega_i}. The zero-temperature update rule for $m(\bm\sigma)<0$ thus amounts to
\begin{equation}
	\sigma_i = \begin{cases}
	\sgn h_i,&\text{if }\ \sigma_i=-1,\\
	\sigma_i,&\text{if }\ \sigma_i=+1.\end{cases}
\end{equation}
Therefore, inactive units are updated based on their field $h_i$, until the overall activity vanishes, \textit{i.e.} $m=0$. This fundamentally distinguishes our kinetic encoding model from the case of energetic encoding described above, in which both active and inactive units are updated at zero temperature [Eq.~\eqref{eq:update_ener}]. As emphasized in the text, since the stability criterion for energetic depends only on the field $h_i$ acting on unit~$i$, it applies also to kinetic encoding.

\begin{figure}[!h]
	\vspace{2mm}
    \centerline{\includegraphics[width=.73\linewidth]{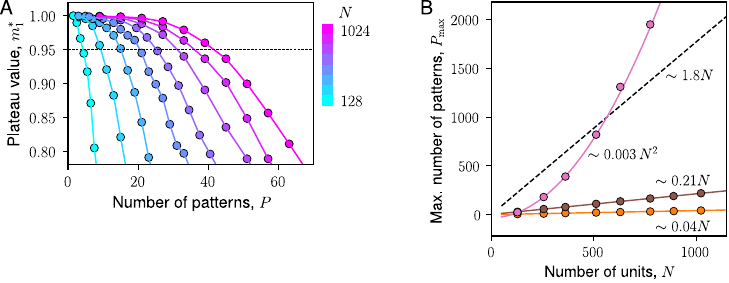}}
   \caption{\label{fig:memory_kin_all} {\it Encoding capacity for kinetic encoding.} {\bf A.} The plateau value of the overlap with the cued pattern, $m_1^*$, decreases with increasing number of patterns, $P$. We define the encoding capacity $P_{\rm max}$ as the value of $P$ at which $m_1^*=0.95$ (dashed line). Here, the initial state has $m_1(0)=0.2$. {\bf B.} We report the values of $P_{\rm max}$ for increasing system size $N$ from {\bf A} here (orange). For Hebbian pair couplings, the encoding capacity $P_{\rm max}$ scales with $N$. The encoding capacity increases if the initial state has $m_1(0)=0.9$ (brown). The Gardner bound [Eq.~\eqref{eq:alpha_G}] is shown as a dashed line. By contrast, for three-body couplings, $P_{\rm max}$ scales with $N^2$ (pink). This is a reproduction of Fig.~\ref{fig:stability}C of the main text.}
\end{figure}

To test our predictions, we numerically quantify the maximum number of patterns $P_{\rm max}$ retrievable by initializing the system with $m_1(0)=0.2$, $m(0)=-0.8$, and $K=Q=10$ (low temperature). As shown in Fig.~\ref{fig:memory_kin_all}A, the plateau overlap $m_1^*$ decreases with $P$. Defining capacity as the value of $P$ for which $m_1^*=0.95$, we obtain $P_\text{max}\approx0.04N$, in agreement with the threshold for global minima in energetic encoding; see Fig.~\ref{fig:memory_kin_all}B. For initial overlap $m_1(0)=0.9$, patterns can be retrieved even if they are only local minima, yielding a larger capacity $P_\text{max}\approx0.21N$, again comparable to energetic encoding. In both cases, the Gardner bound far exceeds the Hebbian result. Finally, introducing three-body couplings produces the expected $P_\text{max}\sim N^2$ scaling (Fig.~\ref{fig:memory_kin_all}B).  
\newline 

In summary, kinetic encoding achieves storage capacities comparable to energetic encoding, both for pairwise couplings as in the original Hopfield model, and for 3-body interactions as in modern versions, surpassing the Gardner bound. The latter dense associative memory models are of great current interest, underscoring the relevance of our work to the contemporary literature. 

\section{Theory for pattern retrieval and escape in kinetic encoding}\label{sec:theory} 
We hereby set out to properly describe the states and transitions of a network with one kinetically encoded pattern, \textit{i.e.} $P=1$. We then write dynamical equations that we finally solve in the regime of large values of $K,Q$ and $N$. This yields the thresholds $K_{\rm min}$ and $Q_{\rm min}$ for retrieval, the retrieval time $\tau_{\rm ret}$ and lifetime $\tau_{\rm life}$, and the limit trajectory of $m_1$ and $m$ for $K,Q\to\infty$. 
\newline

{\bf Detailed description of the system.}
For $P=1$, as shown in Fig.~\ref{fig:sketch}D, the units fall into four categories depending on their activity $\sigma_i=\pm1$ and target activity $\xi^1_i=\pm1$. We annotate them using green plus and minus signs if they respect the (green) pattern, and with black symbols otherwise. The corresponding fractions of units written $f_\gp,f_\gm,f_+$ and $f_-$ satisfy the relationships $f_\gp+f_-=f_\gm+f_+=\tfrac12$, since the target activities $\pm1$ are equiprobable. We thus further describe any state of the network, $\bm\sigma=(\sigma_1,\dots,\sigma_N)$, solely in terms of the error fractions $f_-$ and $f_+$, from which the activity and overlap result as
\begin{align}\label{eq:mm1vf}
	m =2\big(f_+-f_-\big) \qq{and} m_1 = 1-2\big(f_++f_-\big).
\end{align}
In particular, if $m=m_1-1$, then $f_+=0$, such that there are only $-1$ errors and no $+1$ errors.

\begin{figure}[!h]
    \centerline{\includegraphics[]{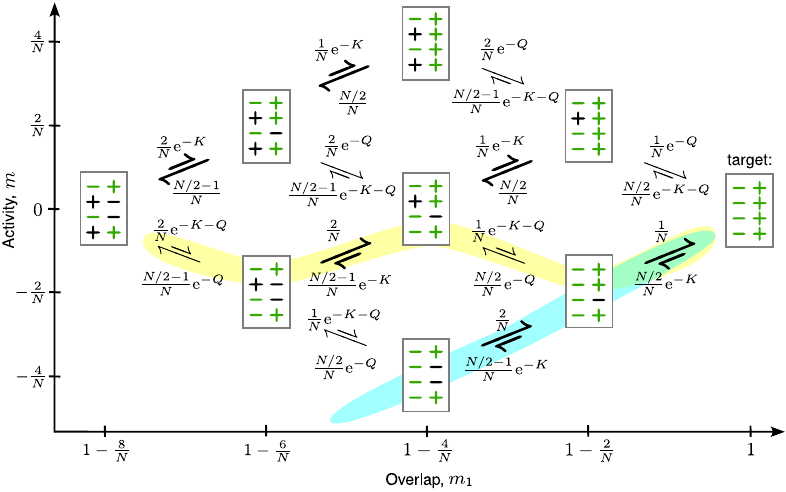}}
   \caption{\label{fig:rate_diag}{\it Kinetic diagram of the retrieval and escape dynamics.} Here, the transition rates are taken from Eq.~\eqref{eq:k-+} in the limit $K,Q\to\infty$ where $\gamma\to1$, with entropic factors taking into account the interchangeability of units that give the same values of $m$ and $m_1$. The limit trajectory for $K,Q\to\infty$ roughly consists of following first the blue path and then the yellow path. On the blue retrieval path, the overlap $m_1=1-2f_-$ increases via $\nearrow$ transitions with rate given by $f_-$, the fraction of $-1$ errors (black minus signs). The yellow escape path consists of double errors (a black $+$ and a black $-$) with rates $\sim\exp(-K-Q)$. The $-1$ error in the state $(m_1=1-\frac4N,m=0)$ can move around on timescale $\tau_m\sim \exp(K)$, by passing through the states $(m_1=1-\frac6N,m=-\frac2N)$ or $(m_1=1-\frac2N,m=\frac2N)$.}
\end{figure}

With this notation, we can rewrite the Glauber transition rates $k(\bm\sigma\to\bm\sigma')$ from Eqs.~(\ref{eq:k_Glauber}-\ref{eq:omega_i}) into four expressions depending on the type of event: correction/creation of a $-1$ error, $k_{-\to\gp},k_{\gp\to-}$, and correction/creation of a $+1$ error, $k_{+\to\gm},k_{\gm\to+}$. We introduce $\gamma=(1+{\rm e}^{-K})^{-1}$ and consider significant overlap values, $m_1\gg1/N$, such that the local field can be approximated as $h_i=\xi_i^1m_1-\sigma_i/N\approx \xi_i^1m_1$. The rates then depend on the activity $m$ as
\begin{equation}\label{eq:k-+}
	\left\{\begin{aligned}	
	k_{-\to\gp}&=\gamma,\qquad\quad k_{\gp\to-}=\gamma\,{\rm e}^{-K},\quad k_{\gm\to+}=\gamma\,{\rm e}^{-Q},\ \ \ \ \quad k_{+\to\gm}=\gamma\,{\rm e}^{-K-Q}, \qq{for} m<0,\\
	k_{-\to\gp}&=\gamma\,{\rm e}^{-K},\quad k_{\gp\to-}=\gamma\,{\rm e}^{-K},\quad k_{\gm\to+}=\gamma\,{\rm e}^{-K-Q},\quad k_{+\to\gm}=\gamma\,{\rm e}^{-K-Q}, \qq{for} m=0,\\
	k_{-\to\gp}&=\gamma\,{\rm e}^{-K},\quad k_{\gp\to-}=\gamma,\qquad\quad k_{\gm\to+}=\gamma\,{\rm e}^{-K-Q},\quad k_{+\to\gm}=\gamma\,{\rm e}^{-Q},\hspace{5mm}\qq{for} m>0. \end{aligned}\right.
\end{equation}  
Here, we see that for large $K$ and $Q$, $k_{+\to\gm}$ vanishes regardless of the value of $m$, making the system inefficient at correcting $+1$ errors. The transition rates are sketched in Fig.~\ref{fig:sketch}D for $m<0$.
Given the interchangeability of units with the same activity and target activity, these transition rates must be multiplied by entropic factors to properly describe the dynamics of the error fractions. This can be visualized in the transition diagram sketched in Fig.~\ref{fig:rate_diag}.
\newline

{\bf Dynamical equations for the error fractions.} 
Now that the system is properly defined, we can investigate its dynamical evolution. As the network is updated spin by spin, the time $t$ measured in number of network updates increases by steps of size $\delta t=1/N$. The error fractions evolve stochastically between time intervals with steps of size $\delta\!f=1/N$. As long as $m =2(f_+-f_-)<0$, given the rates in Eq.~\eqref{eq:k-+}, $f_\pm$ follow two non-coupled random walks with drift, $f_\pm(t+\delta t)-f_\pm(t)=\eta_\pm(t)\delta\!f$, where the noises, $\eta_\pm$, are given by
\begin{align}\label{eq:f+-_sto_m<0}
	\eta_+=\left\{\begin{aligned}
	1,& \qq{with prob.} \gamma\,{\rm e}^{-Q}\big(\tfrac12-f_+\big),\\\
	-1,& \qq{with prob.} \gamma\,{\rm e}^{-K-Q}f_+,
\end{aligned}\right.  \qq{and} \eta_-=\left\{\begin{aligned}
	1,& \qq{with prob.} \gamma\,{\rm e}^{-K}\big(\tfrac12-f_-\big),\\\
	-1,& \qq{with prob.} \gamma f_-.
\end{aligned}\right.
\end{align}
In the thermodynamic limit, $N\gg1$, more analytically tractable, the averages of $f_\pm$ over many realizations (which we also denote by $f_\pm$) will thus satisfy two simple linear differential equations written
\begin{align}\label{eq:df+-_m<0}
	\frac{\text df_+}{\text dt} =\gamma\,{\rm e}^{-Q}\Big[\big(\tfrac12-f_+\big)-{\rm e}^{-K}f_+\Big], \qq{and}
	\frac{\text df_-}{\text dt} =\gamma\Big[{\rm e}^{-K}\big(\tfrac12-f_-\big)-f_-\Big].
\end{align}

We moreover consider the regime of large energetic drive and discrimination barrier, $K,Q\to\infty$, which we expect to yield fast and accurate retrieval of the pattern. In this regime, as $\gamma\to1$ and $\exp(-Q)\to0$, there is a large separation of timescales between the temporal variations of the error fractions $f_-$ and $f_+$. Thus, at first, $f_+$ remains approximately constant while $f_-$ decreases toward the value $\frac12\gamma\,{\rm e}^{-K}\to0$ at which $\frac{\text d}{\text d t}f_-=0$. However, once $f_-$ reaches $f_+$ and $m=0$, the distribution of the noises, $\eta_\pm$, according to~Eq.~\eqref{eq:k-+} now becomes
\begin{align}\label{eq:f+-_sto_m=0}
	\eta_+=\left\{\begin{aligned}
	1,& \qq{with prob.} \gamma\,{\rm e}^{-K-Q}\big(\tfrac12-f_+\big),\\\
	-1,& \qq{with prob.} \gamma\,{\rm e}^{-K-Q}f_+,
\end{aligned}\right.  \qq{and} \eta_-=\left\{\begin{aligned}
	1,& \qq{with prob.} \gamma\,{\rm e}^{-K}\big(\tfrac12-f_-\big),\\\
	-1,& \qq{with prob.} \gamma\,{\rm e}^{-K} f_-.
\end{aligned}\right.
\end{align}
Since $m=0$ is a stable state of the potential $K|m|$, the fast-varying function $f_-$ will tend to stay close to $f_+$, while the slow-varying function $f_+$ dictates the complex long-time dynamics, with a timescale $\sim\exp(K+Q)$.
\newline

\begin{figure}[!h]
    \centerline{\includegraphics[scale=.75]{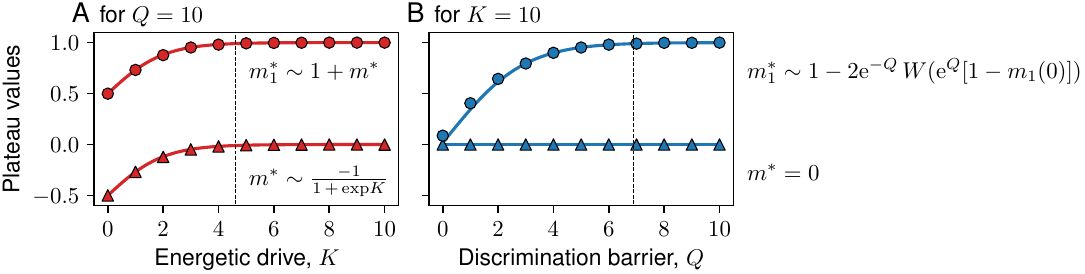}}
   \caption{\label{fig:m1star}{\it Kinetically stable values.} This shows the dependence of the plateau values from trajectories of the type in Fig.~\ref{fig:param_sweep}A-B. {\bf A.} For large $Q$, the overlap $m_1^*$ and activity $m^*$ satisfy Eq.~\eqref{eq:m1*_K}. The dashed line denotes the analytical estimate for $1\%$ error, $K_{\rm min}\approx4.6$. {\bf B.} For large $K$, the overlap $m_1^*$ satisfies Eq.~\eqref{eq:m1*_Q}, while $m^*=0$. The dashed line denotes the analytical estimate for $1\%$ error, $Q_{\rm min}\approx6.9$. Here, as in Fig.~\ref{fig:param_sweep}, $m_1(0)=0.2$ and $m(0)=-0.8$.}
\end{figure}

\begin{figure}[!h]
    \centerline{\includegraphics[scale=1.07]{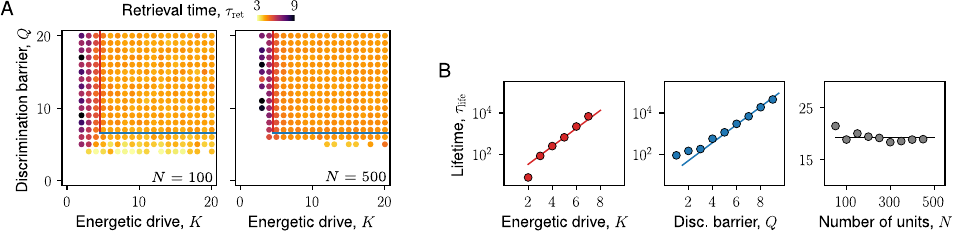}}
   \caption{\label{fig:tau_ret_life}{\it Characteristic timescales verify predictions.} {\bf A.} The retrieval time, estimated in retrieval trajectories as the time at which $m_1=0.99$, satisfies the analytical estimate, $\tau_{\rm ret}\approx 4.4$, within the retrieval region (inside the blue and red lines). Here, $P=1$, $m_1(0)=0.2$ and $m(0)=0.8$. {\bf B.} The lifetime, estimated in escape trajectories as the time at which $m_1=0.8$, satisfies the analytical estimate, $\tau_{\rm life}\sim \exp(K+Q)$. Here, $P=1$, $m_1(0)=1$ and $m(0)=0$, and we vary $K,Q$ or $N$ while keeping the other two constant.}
\end{figure}

{\bf Solutions of the dynamical equations.} 
We now set out to solve the set of dynamical equations derived above for large values of $K,Q$ and $N$. From an initial state with $f_-(0)>f_+(0)$, the average fraction of $+1$ errors, $f_+$, will at first remain close to its initial value, while the average fraction of $-1$ errors, $f_-$, decreases toward its kinetically stable value $f_-^*\approx\max\big\{\frac12\gamma\,{\rm e}^{-K},f_+(0)\big\}$. Both will then monotonically reach their steady-state values $f_+^{\rm ss}=f_-^{\rm ss}=\frac14$. On average, the activity $m$ and overlap $m_1$, defined in Eq.~\eqref{eq:mm1vf}, will thus at first both increase from their initial values to reach kinetically stable values $m^*$ and $m_1^*$. In the limit $K,Q\to\infty$ and $N\gg1$ and for $f_+(0)=0$, these read
\begin{equation}\label{eq:m1*_K}
	m^*\approx-\frac{1}{1+\exp(K)} \qq{and} m_1^*\approx1+m^*, 
\end{equation}
in agreement with the simulated data shown in Fig.~\ref{fig:m1star}A. The minimum value of $K$ needed to obtain only $1\%$ error ($m_1=0.99$) is thus $K_{\rm min}\approx4.6$. During retrieval, $m_1(t)=1-2f_-(t)$, where Eq.~\eqref{eq:df+-_m<0} prescribes that $f_-(t)\approx f_-(0)\exp(-t)$. The retrieval time to reach $m_1=0.99$ from $m_1(0)=0.2$ thus reads $\tau_{\rm ret}\approx4.4$, in agreement with the data in Fig.~\ref{fig:tau_ret_life}A. Moreover, the activity and overlap are related as 
\begin{equation}\label{eq:m1tvmt}
	m_1(t)\approx 1+m(t)
\end{equation}
during the whole retrieval phase; see the blue path in Fig.~\ref{fig:rate_diag}. While the steady-state activity is already attained, $m^{\rm ss}=m^*$, the overlap $m_1$ will then monotonically reach its steady-state value $m_1^{\rm ss}=0$. The escape of $m_1$ from the kinetic trap (yellow path in Fig.~\ref{fig:rate_diag}) occurs on a timescale corresponding to the pattern lifetime, which we estimate as the rate to make $\sim N$ double errors, $\tau_{\rm life}\sim \exp(K+Q)$, in agreement with the data in Fig.~\ref{fig:tau_ret_life}B. This yields Eq.~\eqref{eq:nretlife} in the main text.

To obtain the dependence of $m_1^*$ on $Q$ and thus the threshold $Q_{\rm min}$, we consider the limit $K\to\infty$. For $f_+(0)=0$, by solving Eq.~\eqref{eq:df+-_m<0} we find that the average fractions evolve as 
\begin{equation}
	f_+(t)\approx\frac12\big(1-{\rm e}^{-\exp(-Q)t}\big)\approx \frac12{\rm e}^{-Q}t \qq{and} f_-(t)\approx f_-(0){\rm e}^{-t}
\end{equation}
until they meet at a time which in the limit $Q\to\infty$ approaches $W\big(2{\rm e}^{Q}f_-(0)\big)$, where $W$ is the Lambert function. This defines the kinetically stable state with activity $m^*=2(f_+^*-f_-^*)=0$ and overlap $m_1^*=1-2(f_+^*+f_-^*)$, while $m_1(0)=1-2f_-(0)$, \textit{i.e.}
\begin{equation}\label{eq:m1*_Q}
	m_1^*\approx 1-2{\rm e}^{-Q}W\big({\rm e}^{Q}[1-m_1(0)]\big).
\end{equation}
This is in agreement with the data in Fig.~\ref{fig:m1star}B. The minimum value of $Q$ to obtain only $1\%$ error from $m_1(0)=0.2$ is thus $Q_{\rm min}\approx6.9$.

\begin{figure}[!h]
    \centerline{\includegraphics[]{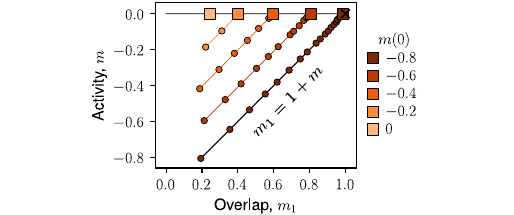}}
   \caption{\label{fig:minit}{\it Limit trajectories from initial states with different activities.} The trajectories in the limit $K,Q\to\infty$ are linear and satisfy Eq.~\eqref{eq:m1(t)} until the plateau with values $m^*=0$ and $m_1^*=m_1(0)-m(0)$ (larger filled markers). Here, all trajectories have $K=Q=6$, $N=2^{10}$, $P=1$ and $m_1(0)=0.2$.}
\end{figure}

Additionally, in the presence of $+1$ errors with $f_+(0)\gg \frac12(1+{\rm e}^K)^{-1}$ and for $K,Q\to\infty$, we instead find $f_-^*=f_+^*=f_+(0)$ and thus
\begin{equation}\label{eq:m1*_f+init}
	m^*=0, \qq{and} m_1^*\approx1-4f_+(0)\approx m_1(0)-m(0).
\end{equation}
Moreover, since Eq.~\eqref{eq:df+-_m<0} prescribes $\frac{\text d}{\text dt}f_- \approx -f_-\gg \frac{\text d}{\text dt}f_+$, we approximate that the average fraction of $-1$ errors, $f_-$ decreases at constant $f_+$. This is such that the retrieval trajectory satisfies the equation
\begin{equation}\label{eq:m1(t)}
	m_1(t)-m(t)\approx m_1(0)-m(0),
\end{equation}
in agreement with the data in Fig.~\ref{fig:minit}. This concludes our analytical investigation of the retrieval and escape dynamics.

\section{Simulation details}\label{sec:simu_details} 
To model network dynamics, we performed Monte Carlo simulations with the Metropolis algorithm. We choose $P$ random patterns, we set the couplings $J_{ij}$ based on Eq.~\eqref{eq:Jij}, and we set the values of the $N$ units with a random cue of pattern $\mu=1$, such that $m_1(0)=0.2$ and $m(0)=-0.8$. We make sure that these initial values satisfy the condition $m_1(0)=1+m(0)$, necessary to reach the pattern state $(m_1=1,m=0)$; see Fig.~\ref{fig:minit}. Then, at each iteration, we randomly select a unit and change its activity with a probability equal to its Glauber transition rate [Eq.~\eqref{eq:k-+}].

\section{Phase diagram for varying load}\label{sec:phases}
\begin{figure}[!h]
    \centerline{\includegraphics[width=.8\linewidth]{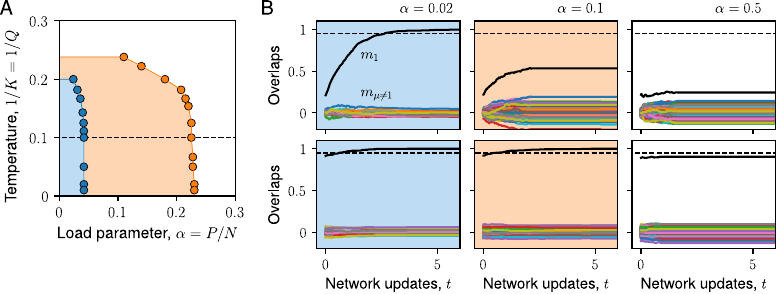}}
   \caption{\label{fig:phase_diagram} {\it Phase diagram for varying load.} {\bf A.} Encoding capacity $P_{\rm max}$ as a function of $K=Q$. The blue curve corresponds to a $5\%$ retrieval error with an initial overlap $m_1(0)=0.2$, while the orange curve corresponds to $m_1(0)=0.9$. The blue and orange regions thus represent global and local stability of patterns, respectively. {\bf B.} Typical trajectories for $K=Q=10$ and varying load $\alpha$, with $m_1(0)=0.2$ (top) and $m_1(0)=0.9$ (bottom). The black dashed line indicates the $5\%$ error threshold in plateau value $m_1^*$. These trajectories are located on the black dashed line in panel {\bf A}.}
\end{figure}

We now investigate another slice of the phase diagram for pattern retrieval. While Fig.~\ref{fig:param_sweep}C examines the case $P=1$ and varying values of $K$ and $Q$, Fig.~\ref{fig:phase_diagram}A explores varying load $\alpha=P/N$ under the constraint $K=Q$. The trajectories in Fig.~\ref{fig:phase_diagram}B reveal three distinct regimes of kinetic pattern stability:
(a) Global stability, where patterns can be retrieved from a distant initial condition (here $m_1(0)=0.2$), for $\alpha \lesssim 0.04$ and $K=Q \gtrsim 5$. 
(b) Local stability, where retrieval is possible only from an initial condition close to the pattern (here $m_1(0)=0.9$), for $\alpha \lesssim 0.2$ and $K=Q \gtrsim 4$.
(c) Instability, where $m_1(t)$ does not increase with time from its initial value.
Interestingly, even at high load $\alpha\sim 1$, the overlap $m_1(t)$ remains close to $m_1(0)$ [Fig.~\ref{fig:phase_diagram}B] up to the lifetime $\tau_{\rm life}\sim\exp(K+Q)$, which diverges at zero temperature. 

The zero-temperature critical values of $\alpha$ are consistent with those shown in Fig.~\ref{fig:memory_kin_all}B and are comparable to those of the classical Hopfield model, as discussed in Sec.~\ref{sec:capa}. The minimum values $K_{\rm min}$ required for retrieval of pattern $\mu=1$ with only $5\%$ error from $m_1(0)=0.2$ or $0.9$ can be estimated from our analytical study [Eqs.~\eqref{eq:m1*_K} and~\eqref{eq:m1*_Q} of Sec.~\ref{sec:theory}]. Since we focus on the case $K=Q$, the relevant threshold is $\min(K_{\rm min},Q_{\rm min})$, giving values $4$ and $4.9$, in agreement with Fig.~\ref{fig:phase_diagram}A. Hence, the phase boundaries between the three retrieval regimes either coincide with those of the Hopfield model or can be quantitatively predicted by our analytical treatment. In particular, regions of global and local stability (blue and orange) are bounded, just as for energetic encoding~\cite{Hertz91,Amit85}.

To analytically estimate the dependence of the prefactor of the capacity scaling $\alpha_{\rm max}$ on $K$ and $Q$, one could apply dynamical mean-field theory~\citep{Sompolinsky88,Clark24}. This represents a significant challenge, which first requires adapting our model to continuous units. Classically, continuous-unit Hopfield networks have been studied at zero temperature~\citep{Hopfield84}. In this limit, our model converges to classical Hopfield networks, and so the dynamical mean-field theory results are equivalent. Therefore, to appreciate the nuances of kinetic encoding for finite $K,Q$, the structure of the thermal noise must be preserved. Although dynamical mean-field theory is well beyond the scope of the present work, we present in the next section an extension of the model to continuous units with appropriate thermal noise.

\section{Variants of the model}\label{sec:variants}
To underscore the relevance of our kinetic encoding framework to neuroscience, we next examine three biologically motivated variants of the model: one with dilute connections (Sec.~\ref{sec:dilute}), another one with sparse patterns (Sec.~\ref{sec:sparse}), and a last one with continuous units (Sec.~\ref{sec:cont}). The first two variants give rise to the capacity scalings shown in Fig.~\ref{fig:dilute_sparse}. 

	\subsection{Variant with dilute connectivity}\label{sec:dilute}
We hereby consider diluting the connections, by randomly pruning a finite fraction of the connections~\cite{Hertz91}. Here, we set $J_{ij}$ to the Hebb value from Eq.~\eqref{eq:Jij} with probability $1-c$ and to $0$ otherwise, while respecting the symmetry $J_{ij} = J_{ji}$. Thus, $c$ is the fraction of diluted bonds. We input these new couplings into the bare kinetic rates $\omega(\bm\sigma,\bm\sigma')$ via the local fields, $h_i=\sum_j J_{ij}\sigma_j$. As for energetic encoding and unsurprisingly, the capacity $P_{\rm max}$ decreases with increasing dilution; see Fig.~\ref{fig:dilute_sparse}A. Therein, $N=2^{10}$, $m_1(0)=0.2$, $K=Q=10$ and we allow for $5\%$ retrieval error.

The observed dependence $P_{\rm max}\sim1-c$ can be rationalized by analyzing the zero-temperature reachability of pattern $\mu=1$ as follows. For $N\to\infty$ and $m_1\to1$, the local fields take the form
\begin{equation}
	h_i\to \xi_i^1 + \sum_{\mu>1}\xi_i^\mu m_\mu.
\end{equation}
Assuming independence between $\xi_i^\mu$ and $m_\mu$, the variance of the noise term is $(P-1)(1-c)/N$. This yields the condition $P_{\rm max}\sim N(1-c)$, consistent with the data in Fig.~\ref{fig:dilute_sparse}A.

	\subsection{Variant with sparse patterns}\label{sec:sparse}
In the following, we introduce a kinetic encoding scheme for patterns that are sparse in inactive units. We then investigate its ability to retrieve a single pattern and assess its encoding capacity. Briefly, we find that, regardless of the sparsity, the dynamics of the overlap and mean activity collapse during retrieval. Furthermore, in agreement with classical works on energetic encoding~\citep{Tsodyks88,Okada96}, we find that the capacity $P_{\rm max}$ increases with sparsity [Fig.~\ref{fig:dilute_sparse}B].
\newline

{\bf Model definition.} 
We now adapt the Hebbian couplings of Eq.~\eqref{eq:Jij} to the case of sparse patterns~\cite{Tsodyks88,Okada96}. We consider patterns with a fraction $a$ of inactive units, \textit{i.e.} $\xi_i^\mu = -1$. For $a\neq\frac12$, the mean activity is shifted from zero to $M=1-2a$, which becomes significant for sparse patterns with $a\ll\frac12$. To compensate for this bias, we modify the couplings between units as
\begin{equation}\label{eq:Jij_sparse}
	J_{ij} = \frac{1}{N(1-M^2)} \sum_{\mu=1}^{P} \left( \xi_i^\mu - M \right) \left( \xi_j^\mu - M \right),
\end{equation}
and the overlaps as $m_\mu=\sum_i (\xi_i^\mu - M)\sigma_i/N(1-M^2)$. Within our kinetic encoding framework, the non-zero mean activity $M$ is incorporated into the Hamiltonian by rewriting Eq.~\eqref{eq:H} as
\begin{equation}\label{eq:ham_sparse}
	\beta\mathcal H(\bm\sigma)=\frac N2K|m(\bm\sigma)-M|.
\end{equation}
Therefore, the energy minima are located at activity $m=M$, while the modified couplings from Eq.~\eqref{eq:Jij_sparse} act on the bare kinetic rates $\omega(\bm\sigma,\bm\sigma')$ of Eq.~\eqref{eq:omega_i} via the local fields, $h_i(\bm\sigma)=\sum_{j\neq i} J_{ij}\sigma_j+M$, where $M$ biases units toward changing activity. 

To study pattern retrieval, we initialize the network in a state where a random fraction $m_1$ of units matches pattern $\mu=1$, while the rest is inactive ($\sigma_i=-1$). This yields $m=(1+M)(m_1-1)+M<M$; see Eq.~\eqref{eq:m1tvmt_a} below. The Hamiltonian in Eq.~\eqref{eq:ham_sparse} then drives the system toward $m = M$ by activating inactive units indiscriminately. However, as per Eq.~\eqref{eq:omega_i}, units with a positive local field will tend to activate first. Using Eq.~\eqref{eq:Jij} and assuming $m_1\gg P/N$, the local fields approximate to $h_i\approx\xi_i^1m_1+M(1-m_1)$. 
In the following, we take $a\le\frac12$ (hence $M\ge0$) and $1-m_1\ll1$, so the first units to activate are those with $\xi_i^1=+1$, leading in theory to retrieval of pattern $\mu=1$. 
\\

{\bf Retrieval of a single pattern.} 
We first investigate the dynamics of a network which encodes a single sparse pattern ($P=1$) with a fraction $a$ of inactive units. In the following, we derive analytical conditions for retrieval as in the case of non-sparse patterns [Sec.~\ref{sec:theory}]. 

The units indexed $i=1,\dots,N$ fall into four categories depending on their activity $\sigma_i=\pm1$ and target activity $\xi^1_i=\pm1$. We annotate them using green plus and minus signs if they respect the (green) pattern, and with black symbols otherwise. Here, the fractions of units written $f_\gp,f_\gm,f_+$ and $f_-$ satisfy the relationships $f_\gp+f_-=1-a$ and $f_\gm+f_+=a$. The mean activity and overlap result as
\begin{align}\label{eq:mm1vf_a}
	m =M+2\big(f_+-f_-\big) \qq{and} m_1 = \frac{1-2\big(f_++f_-\big)-Mm}{1-M^2}.
\end{align}
With this notation, the Glauber transition rates $k(\bm\sigma\to\bm\sigma')$ have a similar expression than for the case $a=\frac12$ [Eq.~\eqref{eq:k-+}], except that the threshold activity value is not $m=0$ but $m=M$, given the potential in Eq.~\eqref{eq:ham_sparse}. Therefore, as long as $m<M$, the averages of $f_\pm$ over many realizations in the thermodynamic limit, $N\gg1$, follow two simple linear differential equations written
\begin{align}\label{eq:df+-_m<M_a}
	\frac{\text df_+}{\text dt} =\gamma\,{\rm e}^{-Q}\Big[\big(a-f_+\big)-{\rm e}^{-K}f_+\Big] \qq{and}
	\frac{\text df_-}{\text dt} =\gamma\Big[{\rm e}^{-K}\big(1-a-f_-\big)-f_-\Big],
\end{align}
where $\gamma=1/[1+\exp(-K)]$. 

\begin{figure}[!b]
    \centerline{\includegraphics[width=.45\linewidth]{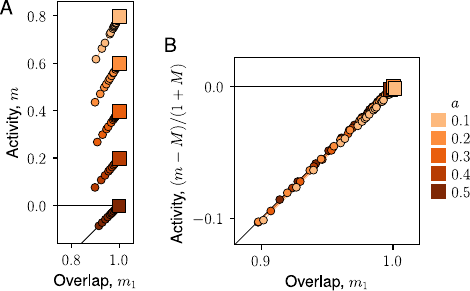}}
   \caption{\label{fig:sparse} {\it Retrieval trajectories and encoding capacity for sparse patterns.} 
    {\bf A.} Stochastic trajectories for $P=1$ and $N=2^{10}$, starting from a random state with $m_1(0)=0.9$, up to the plateau value $m_1^*\approx1$ (squares). Here $K=Q=10$ and the sparsity values $a=0.1,\dots,0.5$ correspond to $M=0.8,\dots,0$. These trajectories are similar to those of Fig.~\ref{fig:param_sweep}D for large $K$ and $Q$.
	{\bf B.} The trajectories for different values of $a$ (and thus $M=1-2a$) follow Eq.~\eqref{eq:m1tvmt_a}.}
\end{figure}

For $f_+(0)=0$ and in the regime of large energetic drive and discrimination barrier, $K,Q\to\infty$, at first, $f_+$ remains approximately constant, $f_+\approx0$ while $f_-$ decreases toward the value $f_-^*\approx (1-a)\gamma\exp(-K)\to0$ at which $\frac{\text d}{\text d t}f_-=0$. The escape dynamics then occurs on a timescale $\sim\exp(K+Q)$. On average, the activity $m$ and overlap $m_1$ will thus at first both increase from their initial values to reach kinetically stable values 
\begin{equation}\label{eq:m1*_K_a}
	m^*\approx M-\frac{1+M}{1+{\rm e}^{K}} \qq{and} m_1^*\approx1+\frac{m^*-M}{1+M}\approx 1-\frac{1}{1+{\rm e}^{K}}. 
\end{equation}
The minimum value of $K$ needed to obtain only $1\%$ error thus remains $K_{\rm min}\approx4.6$. During retrieval, $m_1(t)=1-f_-(t)/(1-a)$, where Eq.~\eqref{eq:df+-_m<M_a} prescribes that $f_-(t)\approx f_-(0)\exp(-t)$. The retrieval time to reach $m_1=0.99$ from $m_1(0)=0.2$ thus reads again $\tau_{\rm ret}\approx4.4$. Moreover, the activity and overlap are related as 
\begin{equation}\label{eq:m1tvmt_a}
	m_1(t)\approx 1+\frac{m(t)-M}{1+M}
\end{equation}
during the whole retrieval phase; see Fig.~\ref{fig:sparse}A-B. In the $M\to0$ limit, this is consistent with Eq.~\eqref{eq:m1tvmt}.

To obtain the dependence of $m_1^*$ on $Q$ and thus the threshold $Q_{\rm min}$, we consider the limit $K\to\infty$. By solving Eq.~\eqref{eq:df+-_m<M_a} we find that the average fractions evolve as 
\begin{equation}
	f_+(t)\approx a\big(1-{\rm e}^{-\exp(-Q)t}\big)\approx a{\rm e}^{-Q}t \qq{and} f_-(t)\approx f_-(0){\rm e}^{-t}
\end{equation}
until they meet at a time which in the limit $Q\to\infty$ approaches $W\big({\rm e}^{Q}f_-(0)/a\big)$, where $W$ is the Lambert function. This defines the kinetically stable state with activity $m^*=M$ and overlap $m_1^* = 1-2(f_+^*+f_-^*)/(1-M^2)$, while $m_1(0)=1-f_-(0)/(1-a)$, \textit{i.e.} 
\begin{equation}\label{eq:m1*_Q_a}
	m_1^*\approx 1-\frac {{\rm e}^{-Q}}{1-a}W\left({\rm e}^{Q}[1-m_1(0)]\frac{1-a}a\right).
\end{equation}
The minimum value of $Q$ to obtain only $1\%$ error thus varies mildly between $Q_{\rm min}\approx6.6$ and $6.9$ for $a$ between $0.1$ and $0.5$. 

Therefore, our main results regarding threshold $K_{\rm min}$, retrieval time $\tau_{\rm ret}$ remain unchanged for sparse patterns, while $Q_{\rm min}$ changes slightly for the considered values of $a$. As for the limit trajectories of $m_1$ and $m$  as $K,Q\to\infty$ for varying sparsity, we find that they collapse onto a master curve, following Eq.~\eqref{eq:m1tvmt_a} [Fig.~\ref{fig:sparse}B].
\\

{\bf Encoding capacity.}
For energetic encoding with the couplings of Eq.~\eqref{eq:Jij_sparse}, the encoding capacity typically increases as patterns become sparser, \textit{i.e.} as $a$ decreases. In the sparse limit $a\ll\tfrac{1}{2}$ and at zero temperature, introducing optimized biases in the local fields yields an encoding capacity $\alpha_c \approx 1/(2a|\ln a|)$~\cite{Tsodyks88,Okada96}, which diverges as $a\to0$. 

For kinetic encoding, even with the simple bias $M$, we observe the expected increase in the maximal number of storable patterns $P_{\rm max}$ as patterns become sparser in inactive units; see Fig.~\ref{fig:dilute_sparse}B. Therein, $N=2^{7}$, $m_1(0)=0.9$, $K=Q=20$ and we allow for $5\%$ retrieval error. The observed dependence $P_{\rm max}\sim1/a$ can be rationalized by analyzing the zero-temperature reachability of pattern $\mu=1$ as follows. For $N\to\infty$ and $m_1\to1$, the local fields take the form
\begin{equation}
	h_i\to \xi_i^1 + \sum_{\mu>1}(\xi_i^\mu-M)m_\mu.
\end{equation}
Assuming independence between $(\xi_i^\mu - M)$ and $m_\mu$, the variance of the noise term $\sum_{\mu>1}(\xi_i^\mu - M)m_\mu$ is $(P-1)(1-M^2)/N$. This yields the condition $P_{\rm max}\sim N/4a(1-a)$, and thus, in the limit $a\to0$, $P_{\rm max}\sim 1/a$, consistent with the data in Fig.~\ref{fig:dilute_sparse}B.

	\subsection{Variant with continuous units}\label{sec:cont}
In the following, we formulate the dynamics of associative neural networks with graded responses as gradient descent dynamics, with adequate choice of diffusivity and noise. We show that the classical continuous-unit Hopfield model~\citep{Hopfield84} falls into this class of models, and thus can be interpreted as an overdamped relaxation to equilibrium. We then introduce a kinetic encoding scheme and assess its performance by essentially reproducing Figs.~\ref{fig:param_sweep} and~\ref{fig:stability} of the main text for this model.
\newline

{\bf General setting.}
Consider a system with real-valued variables $x_i\in\mathbb R$, $i=1,\ldots, N$ evolving synchronously in an energy landscape $E(\mathbf x)$. In Stratonovich’s convention, the overdamped Langevin dynamics reads as in Eq.~\eqref{eq:dxdt_gen}:
\begin{align}\label{eq:dxdt_gen_2}
    \frac{\text d x_i}{\text d t}=-\mu_i(\mathbf x)\frac{\partial E}{\partial x_i} + \sqrt{2D_i(\mathbf x)}\eta_i(t),
\end{align}
where $\bm\mu$ is a mobility vector, $\mathbf D$ a diffusivity vector, and $\eta_i$ are independent Gaussian white noise terms of zero mean and unit variance. This dynamics corresponds to an equilibrium process if the stationary probability distribution is of Boltzmann type and if no probability currents are present, \textit{i.e.} detailed balance. This requires the fluctuation-dissipation relation $\mathbf D = \bm\mu k_{\rm B}T=\bm\mu/\beta$ and another trickier condition. To specify the latter, we write the Fokker-Planck equation for the probability density $P(\mathbf x,t)$ with current $J_i=(-\mu_i\partial_i E+\frac12\partial_i D_i)P-\partial_i(D_iP)$ as $\partial_t P=-\sum_i\partial_i J_i$~\cite{van_Kampen07}. Assuming an equilibrium distribution $P_{\rm eq}\propto\exp(-\Phi(\mathbf x))$, the zero-current condition $\bf J=0$ yields $\partial_i\Phi=\beta \partial_i E+\frac12\partial_i\ln \mu_i$. For a scalar potential $\Phi$ to exist, the mixed derivatives must commute, leading to the zero-curl condition $\partial_j\partial_i\ln\mu_i=\partial_i\partial_j\ln\mu_j$. In this case, the equilibrium distribution reads $P_{\rm eq}\propto\prod_i\sqrt{1/\mu_i(\mathbf x)}\exp(-\beta E(\mathbf x))$.
\newline

{\bf Energetic encoding.}
In Hopfield's original formulation, $x_i$ denotes the input to unit $i$, and its output $y_i$ is determined by a sigmoidal function $g:\mathbb R\to(-1,1)$ with gain $\lambda$. This is such that, at high gain, $\lambda\to\infty$, the outputs $y_i=g(x_i)$ approach the binary variables $\sigma_i\in\{-1,1\}$ of the discrete case. We further denote the inverse function of $g$ as $g^{-1}$ and its derivative as $g'$. The network energy is defined over the outputs $y_i=g(x_i)$ as~\citep{Hopfield84,Krotov21} 
\begin{equation}\label{eq:E(y)_Jij}
	E(\mathbf x)=-\frac12\sum_{i,j\neq i}J_{ij}y_iy_j + \sum_i\int_0^{y_i}g^{-1}(y)\text dy,
\end{equation}
where pattern states are stored in the couplings $J_{ij}$ as in Eq.~\eqref{eq:Jij}. Choosing the mobilities as $\mu_i=1/g'(x_i)$ ensures detailed balance (since $\partial_j\partial_i\ln g'(x_i)=0$) and yield the drift term $-\mu_i\frac{\partial E}{\partial x_i}=-\frac{\partial E}{\partial y_i}$. From the energy in Eq.~\eqref{eq:E(y)_Jij}, the relaxation dynamics follows as
\begin{equation}\label{eq:dxdt_ener}
	\frac{\text dx_i}{\text dt}=\sum_{j\neq i}J_{ij}g(x_j)-x_i+\sqrt{\frac{2k_{\rm B}T}{g'(x_i)}}\eta_i(t), 
\end{equation}  
which reduces, in the zero temperature limit ($T=0$), to the deterministic dynamics of Ref.~\citep{Hopfield84}. Therefore, the general formalism in Eq.~\eqref{eq:dxdt_gen} accounts for the equilibrium dynamics of standard Hopfield networks with continuous units. 

To clarify the notion of equilibrium, we now examine the probability distribution. For $\lambda\to\infty$, as $g$ approaches the sign function, $g'$ becomes constant (except at $x=0$), and the second term in Eq.~\eqref{eq:E(y)_Jij}, representing the area under the inverse, vanishes. Hence, the equilibrium distribution defined on the binary outputs $\sigma_i$ reduces to $P_{\rm eq}\propto\exp(-\beta\mathcal H(\bm\sigma))$, where the constant factor $\prod_i\sqrt{g'(x_i)}$ has been absorbed into the proportionality constant, and $\mathcal H=-\frac12\sum_{i,j\neq i}J_{ij}\sigma_i\sigma_j$ is the Hamiltonian of the discrete case. This establishes consistency between Eq.~\eqref{eq:k_Glauber} for discrete units and Eq.~\eqref{eq:dxdt_gen} or~\eqref{eq:dxdt_gen_2} for continuous units.
\newline

{\bf Kinetic encoding.}
We now construct a kinetic encoding scheme for networks of continuous units, in direct analogy to the discrete case. In particular, we propose an energy function that does not depend on the couplings $J_{ij}$, but rather on the mean network activity $m=\sum_ig(x_i)/N$ as in Eq.~\eqref{eq:E(x)_kin}:
\begin{equation}\label{eq:E(x)_kin_2}
	E(\mathbf x)=\frac K\lambda|m(\mathbf x)| + \sum_i\int_0^{y_i}g^{-1}(y)\text dy,
\end{equation}
where $K\ge0$ is the energetic drive. Pattern states are instead encoded through unit-specific frequencies $\omega_i(\mathbf x)$ that mimic the bare kinetic rates of Eq.~\eqref{eq:omega_i}:
\begin{equation}\label{eq:omega_i_cont_2}
\omega_i(\mathbf x)=\begin{cases}
	1,&\text{if }\ h_i\ge 0,\\
	\exp(-Q),&\text{if }\ h_i< 0,\end{cases}
\end{equation}
where $Q\ge 0$ is a discrimination barrier and $h_i(\mathbf x)=\sum_{j\neq i}J_{ij}g(x_j)$ is the effective field acting on unit $i$. We incorporate these frequencies into the mobilities as $\mu_i=\omega_i(\mathbf x)/g'(x_i)$, which satisfies detailed balance. (Indeed $\partial_j\partial_i\ln\mu_i=\partial_j\partial_i\ln\omega_i=0$, because $\omega_i$ is function of $h_i$ which depends on all $x_{k\neq i}$.) The pattern information in $J_{ij}$ thus affects only the kinetics. The resulting equilibrium dynamics is as in Eq.~\eqref{eq:dxdt_kin}:
\begin{equation}\label{eq:dxdt_kin_2}
	\frac{\text dx_i}{\text dt}=\omega_i(\mathbf x)\left[-\frac K\lambda\sgn(m(\mathbf x))-x_i\right]+\sqrt{\frac{2k_{\rm B}T\omega_i(\mathbf x)}{g'(x_i)}}\eta_i(t),
\end{equation}
which constitutes another instance of the overdamped relaxation in Eq.~\eqref{eq:dxdt_gen}. 

For the equilibrium distribution, an additional factor appears: $\prod_i\sqrt{1/\omega_i}=\exp(N_h^-Q/2)$, where $N_h^-$ is the number of negative $h_i$. To evaluate $N_h^-$, we sum the $h_i$ given the $J_{ij}$ of Eq.~\eqref{eq:Jij}: We find $\sum_i h_i\propto\sum_i\xi_i^\mu=0$, since the patterns have null mean. Consequently, regardless of the network state $\bf x$, exactly half of the $h_i(\mathbf x)$ are negative, $N_h^-=N/2$, allowing the product to be absorbed into the proportionality constant. Thus, for $\lambda\to\infty$, the equilibrium distribution reduces to $P_{\rm eq}\propto\exp(-\beta\mathcal H(\sigma))$, with $\mathcal H$ as in Eq.~\eqref{eq:H}, showing consistency in the limit of binary outputs. NB: the continuous energetic drive $K^\text{cont}$ relates to its discrete counterpart via $\beta K^\text{cont}/\lambda=K^\text{disc}N/2$. In what follows, we investigate Eq.~\eqref{eq:dxdt_kin_2} numerically at zero temperature ($T=0$), using the input-output function $g(x)=\tanh(\lambda x)$.
\newline

{\bf Numerical results.}
To probe the performance of the kinetic encoding scheme in Eqs.~(\ref{eq:omega_i_cont_2}-\ref{eq:dxdt_kin_2}), we begin with the study of the retrieval of a single pattern, $P=1$. We consider synchronous network updates over time $t$ corresponding to the number of updates multiplied by the time increment ($\delta t=0.005$). We set the gain to a high value ($\lambda=5$) and investigate the dynamics of overlaps $m_\mu=\sum_ig(x_i)\xi_i^\mu/N$ for varying values of $K$ and $Q$. As in the discrete case, we initialize the network with a cue of pattern $\mu=1$, and the rest inactive, such that $m_1>0$ and $m=m_1-1<0$. 

\begin{figure}[!h]
    \centerline{\includegraphics[width=.5\linewidth]{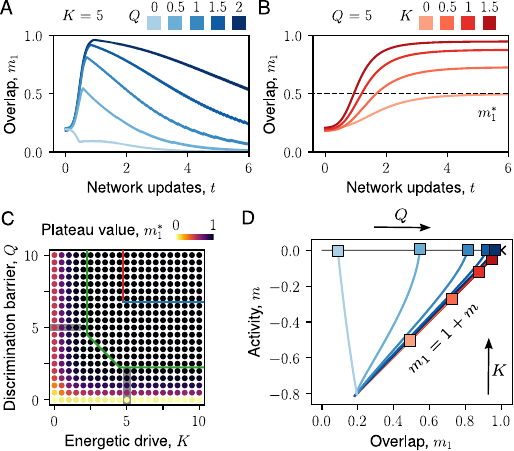}}
   \caption{\label{fig:param_sweep_cont} {\it Successful retrieval of a single pattern with continuous units.} {\bf A.} Stochastic trajectories for $P=1$ and $N=2^{10}$, starting from a random state with $m_1=0.2$ and $m=-0.8$. At a fixed value $K=5$, $m_1$ reaches a maximal value $m_1^*$ which increases with $Q$. {\bf B.} At $Q=5$, $m_1$ displays a plateau with value $m_1^*$ increasing with $K$. {\bf C.} Phase diagram of pattern retrieval. The plateau value $m_1^*$ corresponds to a $1\%$ error for $Q$ and $K$ above the green curve. The grey regions refer to the time trajectories shown in {\bf A-B}. The blue and red lines denote the thresholds of the discrete case [Fig.~\ref{fig:param_sweep}C]. {\bf D.} The blue trajectories from {\bf A} have a short-lived plateau value $m^*\approx0$ (larger squares), while for the red trajectories from {\bf B}, $m_1^*\approx1+m^*$.}
\end{figure}

Figure~\ref{fig:param_sweep_cont} shows results largely similar to Fig.~\ref{fig:param_sweep}. Specifically, the overlap $m_1$ transiently plateaus at large values of $K$ and $Q$. Moreover, for $K,Q\to\infty$, the plateau value $m_1^*$ approaches 1 and the dynamics approaches $m_1(t)=1+m(t)$. However, there is one important difference: The separation of timescales required for a long transient plateau is not achieved for low values of $Q$, despite large values of $K$. As shown in Fig.~\ref{fig:param_sweep_cont}A, this is because, contrary to the case of discrete units, in which the lifetime $\tau_{\rm life}$ increases with both $K$ and $Q$; the lifetime increases only with $Q$ for continuous units. Moreover, Fig.~\ref{fig:param_sweep_cont}C shows that the threshold values to reach plateau values $m_1^*\ge0.99$ denoted as $K_{\rm min}$ and $Q_{\rm min}$ are roughly half as small as in the discrete case. 

\begin{figure}[!t]
    \centerline{\includegraphics[width=.5\linewidth]{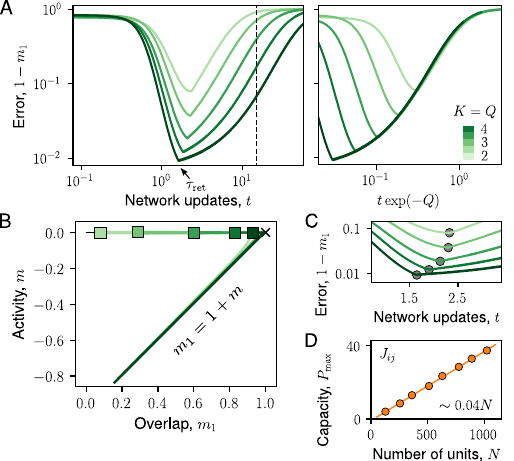}}
   \caption{\label{fig:escape_cont} {\it Kinetic stability and capacity for continuous units.} {\bf A.} The system eventually escapes the kinetic trap ($m_1=1$) toward states with lower values of $m_1$. The escape dynamics defines the lifetime, $\tau_{\rm life}^\text{cont}\sim\exp(Q)$. Here, $P=1$, $N=100$ and $m_1(0)=0.2$. {\bf B.} The trajectories in log time show that the eventual slow decrease of $m_1$ occurs near the $m=0$ line. The values at time $t=15$ (dashed line in {\bf A}) are shown as larger squares. {\bf C.} Both the minimum error and the time of the minimum error decrease with increasing $K=Q$. {\bf D.} The encoding capacity $P_{\rm max}$ at which $m_1^*=0.95$ scales with $N$, with the same prefactor as in the discrete case [Fig.~\ref{fig:memory_kin_all}B].}
\end{figure}

At long times, even for large values of $K$ and $Q$, the network escapes the kinetic trap $m_1^*=1$ to reach lower values of $m_1$ more entropically favorable. Figures~\ref{fig:escape_cont}A-B detail this escape dynamics and show that although the system remains close to the energy minimum $m=0$, the timescale of escape, \emph{i.e.} the lifetime of the pattern is different from the discrete case. As shown in Fig.~\ref{fig:escape_cont}A, we find 
\begin{equation}
	\tau_{\rm life}^\text{cont}\sim\exp(Q),
\end{equation}
which lacks the dependence in $K$ found in the discrete case. Another important difference is that both the minimum error and the time of the minimum error decrease with increasing values of $K=Q$; see Fig.~\ref{fig:escape_cont}C. This corresponds to a counter-intuitive tradeoff typical of kinetic encoding~\citep{Bennett79,Sartori13,Benoist25}. 

For multiple patterns, $P>1$, Fig.~\ref{fig:escape_cont}D shows that the encoding capacity $P_{\rm max}$ scales with $N$ similarly than for the case of discrete units [Fig.~\ref{fig:memory_kin_all}B]. In particular, the prefactor is the same: $P_{\rm max}\approx 0.04 N$ for large $N$, $m_1(0)=0.2$ and 5\% error. 

\begin{figure}[!t]
    \centerline{\includegraphics[scale=.7]{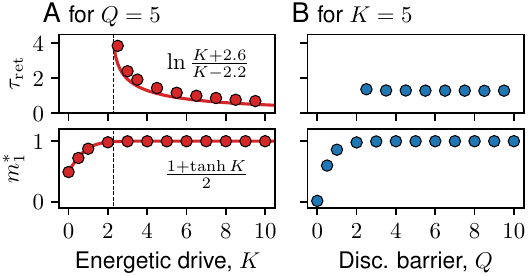}}
   \caption{\label{fig:dep_KQ_cont} {\it Retrieval time and plateau overlap value for continuous units.} We define $\tau_{\rm ret}$ as the time to reach $m_1=0.99$ from $m_1(0)=0.2$. While $m_1^*$ is the plateau value of the overlap with the cued pattern. {\bf A.} We vary $K$ at fixed $Q$. The dashed line denotes the threshold value $K_{\rm min}\approx2.3$ above which $m_1^*\ge0.99$ and $\tau_{\rm ret}$ is well defined. The analytical curves follow Eqs.~\eqref{eq:m1*_cont} and~\eqref{eq:tauret_cont}. Here, $P=1$. {\bf B.} We vary $Q$ at fixed $K$.}
\end{figure}

Additionally, we detail in Fig.~\ref{fig:dep_KQ_cont} the dependence of the retrieval time $\tau_{\rm ret}$ and plateau overlap value $m_1^*$ on the energetic drive $K$ and the discrimination barrier $Q$. As alluded before, $\tau_{\rm ret}$ decreases slightly with increasing $K$, whereas $m_1^*$ shows a qualitatively similar dependence with $K$ and $Q$ than in the discrete case [Fig.~\ref{fig:memory_kin_all}].
\newpage

{\bf Minimal theory.} In the limit $Q\to\infty$ and for a single pattern, we can approximate the retrieval dynamics as follows. Since the local field reads $h_i\approx\xi_i^1m_1$, only units with $\xi_i^1=+1$ are updated, due to their non-zero frequency $\omega_i=1$ [Eq.~\eqref{eq:omega_i_cont_2}]. Since the mean network activity is negative $m<0$, the noiseless dynamics of units with positive target activity is written [Eq.~\eqref{eq:dxdt_kin_2}]
\begin{equation}\label{eq:xit}
	x_i(t)=\frac K\lambda+\left[x_i(0)-\frac K\lambda\right]{\rm e}^{-t}.
\end{equation}
Among these units, a fraction $f_-(0)=(1-m_1(0))/2$ has initial output $g^-\approx-0.99$ (the $-1$ errors) and a fraction $\frac12-f_-(0)=m_1(0)/2$ has initial output $g^+\approx0.99$; see classification of units in Sec.~\ref{sec:theory}. Therefore, the overlap dynamics is
\begin{equation}\label{eq:m1t_cont}
	m_1(t)=\frac12 + \frac{1-m_1(0)}2g(x_i^-(t)) + \frac{m_1(0)}2g(x_i^+(t)),
\end{equation}
where $x_i^\pm(t)$ have initial values $\tanh^{-1}(g^\pm)/\lambda$. The prediction from Eq.~\eqref{eq:m1t_cont} is indistinguishable from the data in Fig.~\ref{fig:param_sweep_cont}B. Taking the limit $t\to\infty$ of Eq.~\eqref{eq:xit}, we obtain $x_i^*=K/\lambda$, $g(x_i^*)=\tanh(K)$ and thus 
\begin{equation}\label{eq:m1*_cont}
	m_1^*=\frac{1+\tanh(K)}2,
\end{equation}
in agreement with the data in Fig.~\ref{fig:dep_KQ_cont}A. The minimum value of $K$ needed to obtain only $1\%$ retrieval error is thus $K_{\rm min}\approx2.3$, consistent with the green curve in Fig.~\ref{fig:param_sweep_cont}C. Then, the retrieval time $\tau_{\rm ret}$, defined as the time to reach $m_1^{\rm th}=0.99$ from $m_1(0)=0.2$ can be estimated from Eq.~\eqref{eq:m1t_cont} by setting $g(x_i^+(t))\approx1$ as 
\begin{equation}\label{eq:tauret_cont}
	\tau_{\rm ret}\approx\ln(\frac{K+\tanh^{-1}|g^-|}{K-\tanh^{-1}(A)}), \qq{with} A=\frac{2m_1^{\rm th}-1-m_1(0)}{1-m_1(0)},
\end{equation}
consistent with the data in Fig.~\ref{fig:dep_KQ_cont}A.
\newline

In conclusion, this constitutes another successful instance of kinetic encoding of pattern states in a network of continuous units. The retrieval is transient, but the kinetic stability of patterns increases rapidly with $K$ and $Q$; and larger networks can encode more patterns. The origin of the differences between the discrete and continuous cases, however, remains to be understood.

\section{Theory for the ageing relaxation}\label{sec:ageing} 
\begin{figure}[!h]
   \centerline{\includegraphics[]{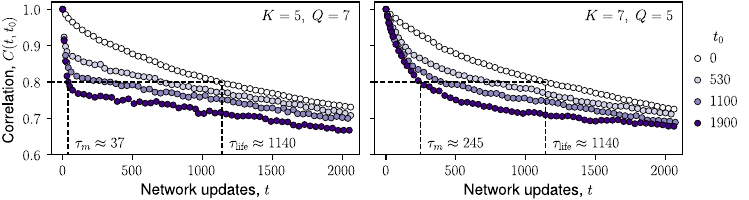}}
   \caption{\label{fig:coor_drop_tau_m}{\it The timescale of the correlation drop can be varied independently of the lifetime.} We define $\tau_{\rm life}$ and $\tau_m$ as the times such that $C(\tau_{\rm life},0)=0.8$ and $C(\tau_m,1900)=0.8$. Like so, we can increase $\tau_m\sim \exp(K)$ at constant $\tau_{\rm life}\sim \exp(K+Q)$ by increasing $K$ while decreasing $Q$. Here, $N=100$, $P=1$, and $m_1(0)=1$, as in Fig.~\ref{fig:coor_drop}.}
\end{figure}

Beyond the decrease in overlap, from $m_1(0)=1$ to $m_1(\tau_{\rm life})=0.8$, we characterize the escape from the pattern via the correlation $C(t,t_0)$ between states at times $t_0$ and $t+t_0$, defined in Eq.~\eqref{eq:Cnn0}. Figure~\ref{fig:coor_drop} shows a clear dependence of $C(t,t_0)$ on the waiting time $t_0$, incompatible with the description of the escape based on a single timescale, \textit{i.e.} the lifetime, $\tau_{\rm life}\sim \exp(K+Q)$. To explain this ageing phenomenon, we introduce another timescale, $\tau_m\sim \exp(K)$, which characterizes the movement of $-1$ errors between units with target activity $+1$; see Fig.~\ref{fig:rate_diag}. For large values of $K$ and $Q$, we thus distinguish three regimes:
\begin{itemize}
	\item (i) On short timescales, $t\ll \tau_m$, $C(t,0)=m_1(t)\approx1$ and $C(t,t_0)\approx1$, whatever $t_0$.
	\item (ii) On intermediate timescales, $\tau_m\lesssim t\ll \tau_{\rm life}$, $C(t,0)\lesssim1$, because relatively few double errors will have occurred. By contrast, for $C(t,t_0)$ with $t_0\gtrsim \tau_{\rm life}$, the many errors in the state $\bm\sigma(t_0)$ with overlap $m_1(t_0)\approx0.8$ will tend to move around in the time window $[t_0,t+t_0]$. This will cause discrepancies with the state $\bm\sigma(t+t_0)$, which explains the important drop in correlation.
	\item (iii) On longer timescales, $t\gtrsim \tau_{\rm life}$, both $C(t,0)$ and $C(t,t_0)$ will decrease significantly with $t$.
\end{itemize}
This explains the absence of time-translation invariance displayed in Fig.~\ref{fig:coor_drop}. The dependence of $\tau_m$ on the energetic drive $K$ and on the discrimination barrier $Q$ is validated in Fig.~\ref{fig:coor_drop_tau_m}.
\newline

\end{document}